\documentclass[12pt]{iopart}
\usepackage{psfig}
\begin{document}

\title{A New Method to Estimate the Noise in
       Financial Correlation Matrices}

\author{Thomas Guhr\dag
        \footnote[3]{To whom correspondence should be 
                     addressed (Thomas.Guhr@matfys.lth.se)}\ 
        and Bernd K\"alber\ddag  
        \footnote[4]{now at:
                    {\it Group Risk Control, 
                         Dresdner Bank AG, 
                         J\"urgen--Ponto--Platz 1, 
                         D--60301 Frankfurt, Germany}}
\address{\dag\ 
         Matematisk Fysik, LTH, Lunds Universitet,
         Box 118, 22100 Lund, Sweden}
\address{\ddag\ 
         Max Planck Institut f\"ur Kernphysik,
         Postfach 103980, 69029 Heidelberg, Germany}
        }

\begin{abstract}
Financial correlation matrices measure the unsystematic correlations
between stocks. Such information is important for risk management.
The correlation matrices are known to be ``noise dressed''. We develop
a new and alternative method to estimate this noise. To this end, we
simulate certain time series and random matrices which can model
financial correlations.  With our approach, different correlation
structures buried under this noise can be detected.  Moreover, we
introduce a measure for the relation between noise and correlations.
Our method is based on a power mapping which efficiently suppresses
the noise.  Neither further data processing nor additional input is
needed.
\end{abstract} 


\submitto{\JPA}

%\maketitle

\section{Introduction}
\label{sec1}

There are different kinds of risk in a stock portfolio, for example, a
systematic one due to the general trends in the economy affecting the
entire market, and an unsystematic one due to events affecting only
segments of the whole market under consideration, such as the
companies within the same industrial branch or in one
country~\cite{PorTh}.  All this is borne out in the time series of the
stocks. In a more general framework, the term ``stock of a company''
should be replaced by ``risk factor''.  The corresponding portfolios
of a bank or an investment company can contain hundreds or thousands
of financial instruments, depending on many different risk
factors~\cite{PorTh}.  However, to be explicit, we will use the term
``stocks'' instead of ``risk factors''. Banks or portfolio managers
are particularly interested in the unsystematic
correlations~\cite{Deutsch,RM}. We refer to them simply as
correlations from now on. The elements of the correlation matrix are
given through the time average over a product of the properly
normalized time series for two stocks.  The number of companies
considered defines the size of this correlation matrix. We will assume
that this number is large enough for a statistical analysis.

In recent years, it was shown that the true correlations one is
interested in can be buried under noise in the commonly used
correlation matrices: Laloux et al.~\cite{Lal} studied an empirical
correlation matrix and found that the bulk part of its spectral
density can be modeled by a purely random matrix. Plerou et
al.~\cite{Pl} worked out spectral fluctuations and also found that
they are compatible with random matrix statistics.  Random matrices
are often used statistical models for quantum chaotic or related
spectral problems, for reviews see Refs.~\cite{Mehta,Haake,GMGW}.  The
presence of random matrix features was coined noise dressing of
financial correlations.  Burda et al.~\cite{Burda} designed a proper
model involving random Levy matrices.  In practice, such noise can
appear if the length of the time series employed to construct the
correlation matrices is too short.  Gopikrishnan et al.~\cite{Gop}
developed a method to reduce the noise by taking advantage of the fact
that large eigenvalues outside the bulk of the spectral density can be
associated with industrial branches. These authors~\cite{Gop} used
this for portfolio optimization, see also Ref.~\cite{Gu2}. The
identification of branches through stock correlations was also studied
by Mantegna~\cite{Man}.

Here, we present a new approach to identify and estimate the noise
and, thereby, also the strength of the true correlations.  As we do
not use the large eigenvalues outside the bulk of the spectral
density, our method is an alternative and a complement to the approach
of Ref.~\cite{Gop}. It will be particularly useful if some of the
large eigenvalues are not large enough to lie outside the bulk of the
spectral density. Our method is based on a power mapping. It does not
require any further data processing or any other type of input. To the
best of our knowledge, it seems to be a new technique for the analysis
of correlation matrices in time series problems.  We have three goals:
first, we develop an alternative method for noise identification,
second, we introduce the power mapping as a new
mathematical--statistical technique and discuss its features in some
detail, third, we present our findings in a self--contained way to
make possible a transfer to other time series problems and the
corresponding correlation matrices.

The article is organized as follows. In Sec.~\ref{sec2}, we outline
the stochastic model, including a review of correlation matrices and a
discussion of the noise dressing phenomenon.  We study the dependence
of the spectral density as a function of the length of the time series
in Sec.~\ref{sec3}.  In Sec.~\ref{sec4}, we introduce the power
mapping and use it to identify and estimate the noise. We summarize
and conclude in Sec.~\ref{sec5}. Three more detailed discussions are
given in the appendix.

\section{Stochastic Model}
\label{sec2}

We formulate the model for financial correlations in Sec.~\ref{sec2.1}.
Since we also aim at a heuristic analytical treatment later on, we
do that in some detail. Correlation matrices and noise dressing
are discussed in Secs.~\ref{sec2.2} and~\ref{sec2.3}, respectively.

\subsection{Normalized Time Series for Correlated Stocks}
\label{sec2.1}

Most models for time series $S(t)$ of stocks~\cite{MS,BP,Voit,PB} have
a random component, involving a random number $\varepsilon$ and a
volatility constant $\sigma^2$, and a non--random component, the drift
part, involving a drift constant $\mu$. Geometric Brownian motion
\begin{eqnarray}
\frac{dS}{S} = \mu dt + \sigma\varepsilon\sqrt{dt} 
\label{eq2.1}
\end{eqnarray}
is particularly popular. The dimensionless quantity $dS/S$ is referred
to as return. Due to the central limit theorem, geometric Brownian
motion leads, largely independently of the distribution for the random
numbers $\varepsilon$, to a log--normal distribution of the stock
prices. This is in fair agreement with the empirical distributions for
price changes about one day and greater~\cite{Stanley1,Stanley2}.  The
tails of the empirical distributions are, in general, much
fatter~\cite{MS,BP}. To describe this, one uses, for example,
autoregressive processes which take into account that the volatility
is also a fluctuating quantity, see the discussion in Ref.~\cite{MS}.
In the present context of correlations, however, it suffices to model
the time series in the spirit of geometric Brownian motion.  In
economics, one wishes to measure the correlations independent of the
drift. Thus, one usually takes the logarithms or the logarithmic
differences of the time series to remove the exponential trend in the
data due to the drift. Moreover, one normalizes the data to zero mean
value and unit variance.

In view of these requirements, Noh~\cite{Noh} suggested a proper model
to study financial correlations.  From an economics viewpoint, it is
of the one--factor type and can be interpreted as an application of
the arbitrage pricing model due to Ross~\cite{Ross}, see also the
discussion in~\ref{app0}. From a physics viewpoint, as pointed out in
Refs.~\cite{Mar,Kull}, Noh's model has much in common with certain
models involving interacting Potts spins~\cite{Wu}. Consider a market
involving $K$ companies, labeled $k=1,\ldots,K$, and $B$ industrial
branches, labeled $b=1,\ldots,B$. The companies within the same
industrial branch are assumed to be correlated. The companies are
ordered in such a way that the indices $k$ of companies within the
same branch follow each other. For example, we have $K=50$ companies
and $B=3$ branches and we assume that the first branch with $b=1$
consists of the first $\kappa_1=5$ companies, the second branch with
$b=2$ consists of the next $\kappa_2=12$ companies, the third branch
with $b=3$ consists of the next $\kappa_3=8$ companies, and, finally,
we assume that the remaining $\kappa=25$ companies are in no
branch. The branch index $b$ is viewed as a function of the company
index $k$, i.e.~we have $b=b(k)$. For the $\kappa$ companies which are
not in any branch, we set $b=b(k)=0$. We refer to $\kappa_b$ as to the
size of the industrial branch $b$. Obviously, we have
\begin{eqnarray}
\sum_{b=1}^B \kappa_b + \kappa = K \ .
\label{eq2.1a}
\end{eqnarray}
Of course, we assume that $\kappa_b>1, \ b=1,\ldots,B$. The number
$\kappa$ can be any non--negative integer, including zero. 

The normalized time series $M_k(t), \ k=1,\ldots,K$ of the returns for
the $K$ companies are modeled as the sum of two purely random
contributions: the first one models the correlations within a given
branch and is thus common to this branch, involving random numbers
$\eta_b(t)$, the second one is specific for the company and involves
random numbers $\varepsilon_k(t)$,
\begin{eqnarray}
M_k(t) = \sqrt{\frac{p_{b(k)}}{1+p_{b(k)}}} \eta_{b(k)}(t)
         +\frac{1}{\sqrt{1+p_{b(k)}}} \varepsilon_k(t) \ .
\label{eq2.2}
\end{eqnarray}
The two contributions are weighted with a parameter $p_{b(k)}$, common
to all companies in the branch $b$.  We assume that the
$\eta_{b(k)}(t)$ and the $\varepsilon_k(t)$ are uncorrelated and
standard normal distributed with zero mean value.  The weights are
assumed to be positive with $p_{b(k)}\ge 0$.  Since the distributions
are symmetric, this is the most general form of the weights.  In the
case that $k$ is not in any branch, i.e.~for $b=0$, we set
$p_{b(k)}=0$.  Here, we use discrete time steps and normalize the time
units such that $dt=1$. The time series $M_k(t)$ consist of $T$ time
values at $t=1,\ldots,T$.

\subsection{Correlation Matrices}
\label{sec2.2}

The time average of a function $F(t)$ over the time series is defined
as
\begin{eqnarray}
\langle F(t) \rangle_T = \frac{1}{T} \sum_{t=1}^T F(t) \ ,
\label{eq2.3}
\end{eqnarray} 
which depends on the length $T$ of the time series.  If the time
series are infinitely long, $T\to\infty$, we have
\begin{eqnarray}
\langle M_k(t) \rangle_\infty = 0
\qquad {\rm and} \qquad
\langle M_k^2(t) \rangle_\infty = 1 \ .
\label{eq2.4}
\end{eqnarray} 
Here, we simply used $\langle \eta_{b(k)}(t)\rangle_\infty=0$,
$\langle \varepsilon_k(t)\rangle_\infty=0$ and 
$\langle\eta_{b(k)}(t)\eta_{b(l)}(t)\rangle_\infty=\delta_{b(k)b(l)}$,
$\langle\eta_{b(k)}(t)\varepsilon_l(t)\rangle_\infty=0$,
$\langle \varepsilon_k(t)\varepsilon_l(t)\rangle_\infty=\delta_{kl}$.

The correlation coefficient between two companies labeled $k$ and $l$
is the average over the product of the two normalized time
series,
\begin{eqnarray}
C_{kl}(T) = \frac{1}{T} \sum_{t=1}^T M_k(t) M_l(t) 
       = \langle M_k(t) M_l(t)\rangle_T \ .
\label{eq2.5}
\end{eqnarray}
If one views the numbers $M_k(t)$ as the entries of a $K\times T$
rectangular matrix $M$, one has
\begin{eqnarray}
C(T) = \frac{1}{T} M M^\dagger 
       = \langle M(t) M^\dagger(t)\rangle_T 
\label{eq2.6}
\end{eqnarray}
for the $K\times K$ correlation matrix $C$.  As these averages depend
on the length $T$ of the time series, we add the argument $T$ to the
correlation matrix. Within our model outlined above, one finds for
infinitely long time series~\cite{Mar}, $T\to\infty$,
\begin{eqnarray}
C_{kl}(\infty) = \lim_{T\to\infty} C_{kl}(T)
               = \frac{1}{1+p_{b(k)}} 
      \left(p_{b(k)}\delta_{b(k)b(l)}+\delta_{kl}\right) \ .
\label{eq2.7}
\end{eqnarray}
Thus, the matrix $C(\infty)$ consists of $B$ square blocks
on the diagonal of dimensions $\kappa_b\times\kappa_b$
with off--diagonal entries $p_{b}/(1+p_{b})$ for branch $b$, and a
$\kappa\times\kappa$ unit matrix for the companies which are in
no branch. The diagonal entries are all unity.  All other entries are
zero: the correlation coefficients between companies which are not in
any branch, those between companies belonging to different branches
and those between a company which is in a branch and another one which
is not. Some issues related to the normalization of correlation
matrices are discussed in~\ref{app0}.

\subsection{Noise Dressing in the Model}
\label{sec2.3}

If the time series have only a finite length, $T<\infty$, it is obvious
that the averages $\langle \eta_{b(k)}(t)\eta_{b(l)}(t)\rangle_T$,
$\langle \eta_{b(k)}(t)\varepsilon_l(t)\rangle_T$ and $\langle
\varepsilon_k(t)\varepsilon_l(t)\rangle_T$ give neither zero nor
unity, but some finite numbers. This describes the noise dressing of
financial correlation matrices which was found in Ref.~\cite{Lal} in
the framework of Noh's model. Due to the finiteness of the time
series, there is a purely random offset to every correlation
coefficient, burying the true correlation coefficient which would be
found if the time series were sufficiently long.  

As we aim at a analytical discussion later on, we formulate the noise
dressing more quantitatively. We employ a standard result of
mathematical statistics~\cite{Krengel}: Consider two time series of
uncorrelated random numbers $\alpha_k(t)$ and $\alpha_l(t),\
t=1,\ldots,T$ with standard normal distributions and zero mean value.
The average $\langle \alpha_k(t)\alpha_l(t)\rangle_T$ is a random
number, following a Gauss distribution centered at mean value unity
with variance $2/T$, if $k=l$, and following a Gauss distribution
centered at mean value zero with variance $1/T$, if $k\ne l$.  Hence,
we can write this average to leading order in $T$ as
\begin{eqnarray}
\langle\alpha_k(t)\alpha_l(t)\rangle_T
       = \delta_{kl} + \sqrt{\frac{1+\delta_{kl}}{T}} a_{kl} \ ,
\label{eq2.8}
\end{eqnarray}
where the $a_{kl}$ are uncorrelated random numbers, independent of
$T$, with standard normal distribution and zero mean value. This
yields for the correlation coefficients the expression
\begin{eqnarray}
C_{kl}(T) &=& \sqrt{\frac{p_{b(k)}}{1+p_{b(k)}}} 
           \sqrt{\frac{p_{b(l)}}{1+p_{b(l)}}} 
           \Biggl(\delta_{b(k)b(l)}
           + \sqrt{\frac{1+\delta_{b(k)b(l)}}{T}} 
                                 a_{b(k)b(l)}\Biggr)
                         \nonumber\\
       & & \quad + \frac{1}{\sqrt{1+p_{b(k)}}}
                   \frac{1}{\sqrt{1+p_{b(l)}}}
           \Biggl(\delta_{kl} 
           + \sqrt{\frac{1+\delta_{kl}}{T}} a_{kl}\Biggr)
                         \nonumber\\
       & & \quad + \sqrt{\frac{p_{b(k)}}{1+p_{b(k)}}}
                         \frac{1}{\sqrt{1+p_{b(l)}}}
                         \frac{1}{\sqrt{T}} a_{b(k)l}
                         \nonumber\\
       & & \quad + \sqrt{\frac{p_{b(l)}}{1+p_{b(l)}}}
                         \frac{1}{\sqrt{1+p_{b(k)}}}
                         \frac{1}{\sqrt{T}} a_{kb(l)} 
\label{eq2.9}
\end{eqnarray}
to leading order in $T$. The first two terms stem from the averages
$\langle \eta_{b(k)}(t)\eta_{b(l)}(t)\rangle_T$ and $\langle
\varepsilon_k(t)\varepsilon_l(t)\rangle_T$, the last two from the
interference averages
$\langle\eta_{b(k)}(t)\varepsilon_l(t)\rangle_T$.  There are four sets
of random numbers, $a_{b(k)b(l)}$, $a_{b(k)l}$, $a_{kb(l)}$ and
$a_{kl}$, respectively.  These sets are understood to be different
from each other.  To avoid a cumbersome notation, we do not add
further indices to specify that. For example, the number
$a_{b(1)b(1)}$ is different from $a_{11}$. As required, the limit
$T\to\infty$ of Eq.~(\ref{eq2.9}) correctly yields Eq.~(\ref{eq2.7}).

\section{Spectral Density and Length of the Time Series}
\label{sec3}

As shown in Ref.~\cite{Lal}, the spectral density is a well suited
observable to study the noise dressing of empirical correlation
matrices.  In Sec.~\ref{sec3.1}, we numerically investigate the
spectral density of our model as a function of the length $T$ of the
time series.  We support our findings by an analytical discussion in
Sec.~\ref{sec3.2}.

\subsection{Numerical Simulation}
\label{sec3.1}

As the number of companies in the analysis~\cite{Lal,Pl} of real
market data is several hundreds, we choose this number as $K=508$ for
our simulation. We assume that there are $B=6$ industrial branches
whose sizes $\kappa_b$ and weights $p_b$ are listed in
table~\ref{tab_NohP2}. Since Noh~\cite{Noh} showed that the choice
\begin{eqnarray}
p_b = 1 - \frac{1}{\kappa_b}
\label{eq3.0}
\end{eqnarray}
leads to spectral densities which are in agreement with the empirical
data~\cite{Lal,Pl}, we use the same parameterization here.  In
addition, we have $\kappa=256$ companies which are not in any branch.

We simulate the correlation matrices $C(T)$ for four different lengths
$T=1650,5000,20000,50000$ of the time series, calculate the
eigenvalues $\lambda_k, \ k=1,\ldots,K$ and work out the spectral
densities $\rho_T(\lambda)$.  The results are shown in
figure~\ref{fig_CEvL_T}. As already observed by Noh~\cite{Noh}, the
first density $\rho_{1650}(\lambda)$ for the shortest time series with
$T=1650$ resembles very much the densities of the empirical
correlation matrices found in Refs.~\cite{Lal,Pl}. There is a bulk
\begin{table}
\caption{\label{tab_NohP2}
         The sizes $\kappa_b$ and the weights $p_b=1-1/\kappa_b$
         for the $B=6$ industrial branches used in the numerical
         simulation.}
\begin{indented}
\lineup
\item[]\begin{tabular}{c|cccccc} 
\br 
$b$          & 1 & 2 &  3 &  4 &  5 &   6 \\ 
\mr
$\kappa_{b}$ & 4 & 8 & 16 & 32 & 64 & 128 \\
$p_b$        & 0.75 & 0.88 & 0.94 & 0.97 & 0.98 & 0.99 \\ 
\br
\end{tabular}
\end{indented}
\end{table}
spectrum in the interval $0<\lambda<2$. Moreover, a number of isolated
peaks in the interval $2<\lambda<70$ is seen. Each of these large
eigenvalues corresponds to an individual industrial branch. For
increasing lengths $T$, the isolated peaks are stable, but the bulk
spectrum separates into two groups of eigenvalues. The left group is
roughly centered around $\lambda=1/2$, the right one around
$\lambda=1$.  We will see in the analytical discussion to be performed
in Sec.~\ref{sec3.2} that the left group can clearly be associated
with the true correlations due to the industrial branches, while the
right group only represents noise around the trivial self--correlation
of the companies for $k=l$.

\subsection{Analytical Discussion}
\label{sec3.2}

\begin{figure}
\begin{minipage}{7.5cm}
\psfig{figure=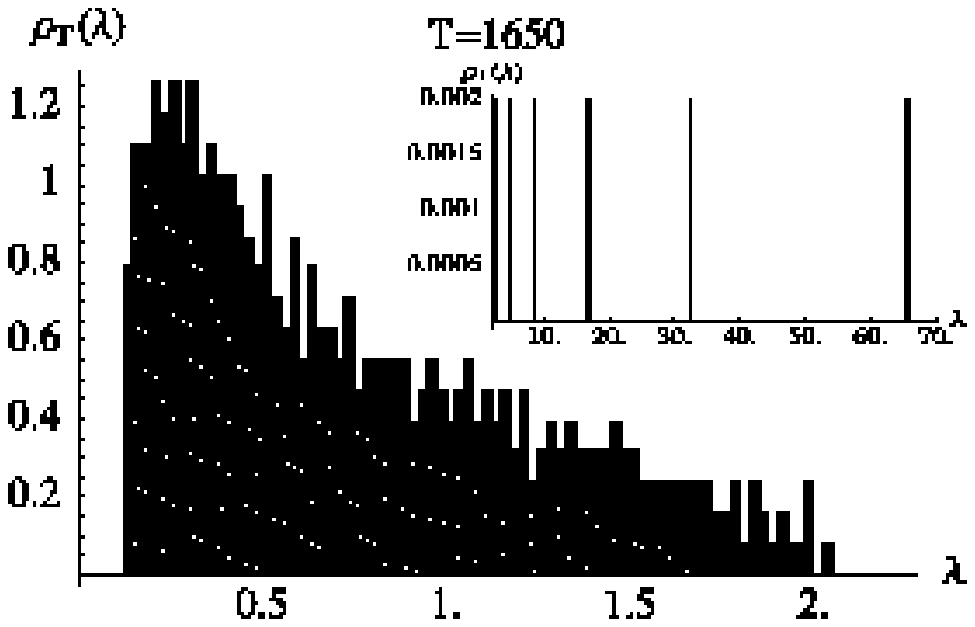,width=7.5cm,angle=0}
\end{minipage}
\hspace{0.2cm}
\begin{minipage}{7.5cm}
\psfig{figure=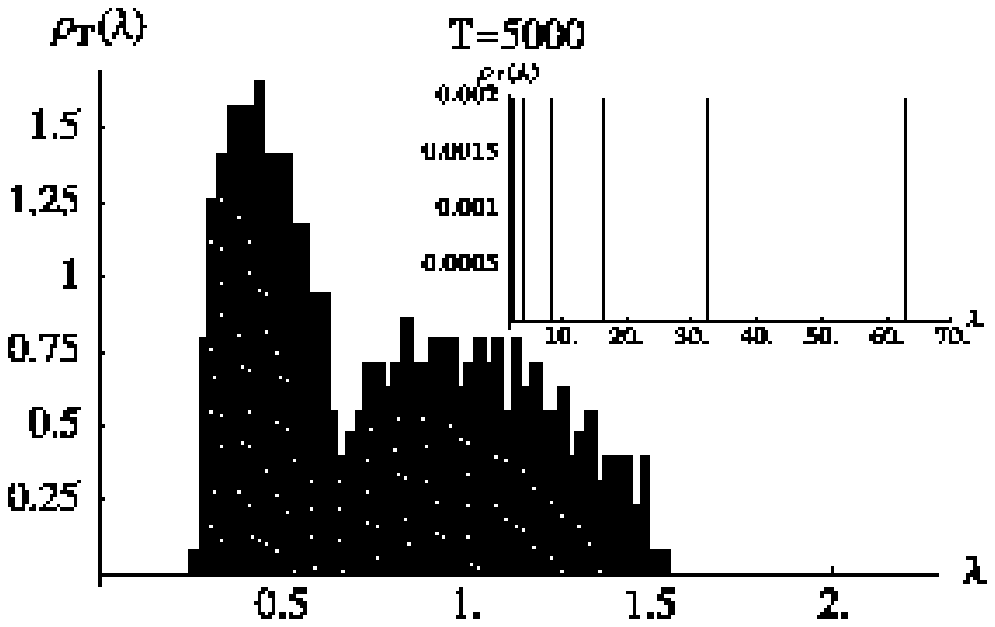,width=7.5cm,angle=0}
\end{minipage}\\ 
\vspace{0.4cm}\\
\begin{minipage}{7.5cm}
\psfig{figure=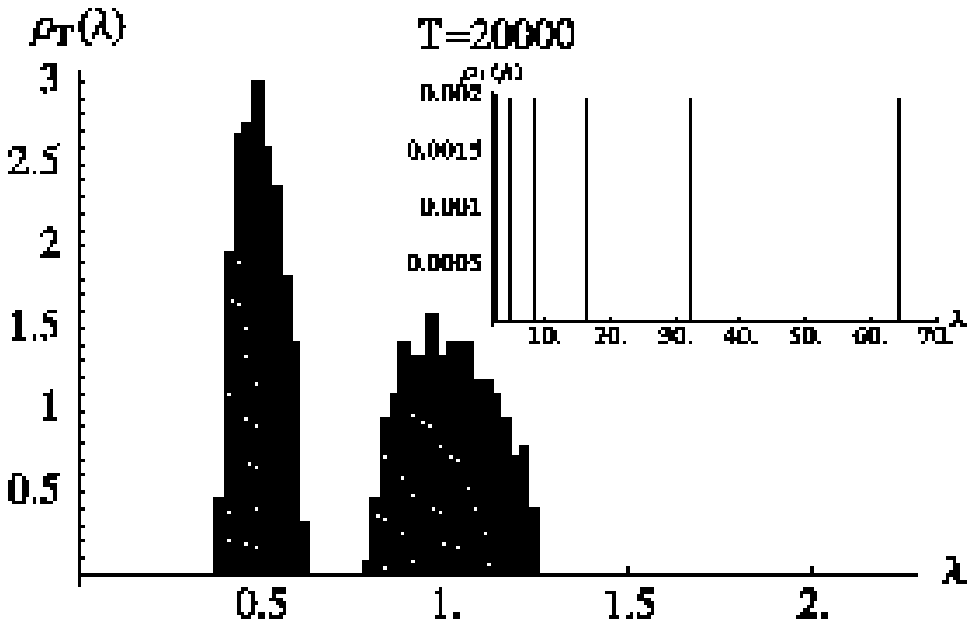,width=7.5cm,angle=0}
\end{minipage}
\hspace{0.4cm}
\begin{minipage}{7.5cm}
\psfig{figure=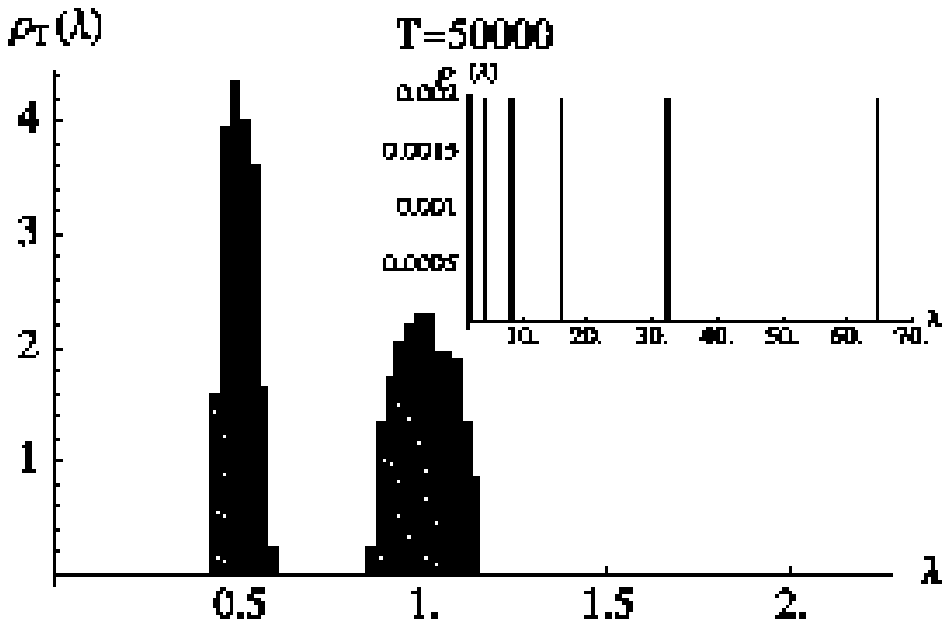,width=7.5cm,angle=0}
\end{minipage}\\
\caption{\label{fig_CEvL_T}
Spectral densities $\rho_T(\lambda)$ of simulated correlation
matrices.  The length $T$ of the time series increases from top to
bottom according to $T=1650,5000,20000,50000$.  The presentation of
every density is split into the regions $0\le\lambda\le 2.2$ and
$2.2\le\lambda\le 70$. The densities are given in units of $K$.}
\end{figure}

For infinitely long time series, $T\to\infty$, the correlation matrix
$C(\infty)$ is block diagonal according to Eq.~(\ref{eq2.7}), and the
spectrum of the eigenvalues $\lambda_k(\infty), \ k=1,\ldots,K$ is
easily calculated. For every branch $b$, we have $\kappa_b-1$
degenerate eigenvalues $1/(1+p_b)$ and one numerically much larger
eigenvalue $(1+\kappa_bp_b)/(1+p_b)$ which is for large size
$\kappa_b$ roughly proportional to $\kappa_b$.  Moreover, there are
$\kappa$ degenerate eigenvalues which are unity for the companies
which are in no branch. Thus, the spectral density is given by
\begin{eqnarray}
\rho_\infty(\lambda)
 &=& \sum_{b=1}^B (\kappa_b-1)\delta\left(\lambda-\frac{1}{1+p_b}\right)
                         \nonumber\\
 & & 
   + \sum_{b=1}^B \delta\left(\lambda-\frac{1+\kappa_bp_b}{1+p_b}\right) 
   + \kappa \delta\left(\lambda-1\right) \ .
\label{eq3.1}
\end{eqnarray} 
This formula helps to understand the last density for the longest time
series with $T=50000$ in figure~\ref{fig_CEvL_T}. The first term
corresponds to the left group, representing the true correlations. The
second term of the $B$ large eigenvalues yield the isolated peaks of
the spectra. The last term in the formula represents the noise peaked
at unity.  The eigenvectors $u_k(\infty), \ k=1,\ldots,K$ can also be
calculated in a straightforward manner for $T\to\infty$.

The density~(\ref{eq3.1}) is smeared out for finite time series,
$T<\infty$.  To estimate the resulting noisy density $\rho_T(\lambda)$
to leading order in $T$, we write the correlation matrix in the form
$C(T)=C(\infty)+C_1/\sqrt{T}$ where $C_1$ can be read off from the
expansion~(\ref{eq2.9}). We also write
$\lambda_k(T)=\lambda_k(\infty)+\lambda_{1,k}/\sqrt{T}$ for the
eigenvalues and $u_k(T)=u_k(\infty)+u_{1,k}/\sqrt{T}$ for the
eigenvectors. One quickly finds
$\lambda_{1,k}=u_k^\dagger(\infty)C_1u_k(\infty)$.  Since the
elements of every $u_k(\infty)$ are numbers depending on the weights
$p_b$ and the sizes $\kappa_b$, every coefficient $\lambda_{1,k}$ is,
according to Eq.~(\ref{eq2.9}), a linear combination of the standard
normal distributed random numbers $a_{b(k)b(l)}$, $a_{b(k)l}$,
$a_{kb(l)}$ and $a_{kl}$. Hence, it is itself a Gaussian distributed
random number $a_k$ with zero mean value and a variance $v_k^2$ which
is some function of the $p_b$ and the $\kappa_b$. We arrive at
\begin{eqnarray}
\lambda_k(T) = \lambda_k(\infty) + \frac{v_k}{\sqrt{T}} a_k
\label{eq3.2}
\end{eqnarray}
for $k=1,\ldots,K$ to leading order in $T$.  We notice that, in this
expansion to order $1/\sqrt{T}$, the Gaussian random variables $a_k$
are uncorrelated.  The effect of the smearing out, i.e.~of the noise
in the model, is marginal for the $B$ numerically large eigenvalues
$(1+\kappa_bp_b)/(1+p_b), \ b=1,\ldots,B$.  Their positions change the
less, the larger the size $\kappa_b$.  The effect of the noise is much
stronger on the numerically smaller eigenvalues.  Moreover, their
degeneracies are lifted and distributions develop around their mean
values.  For the eigenvalues $1/(1+p_b), \ b=1,\ldots,B$ which belong
to the $B$ industrial branches and, hence, describe true correlations,
the mean value is
\begin{eqnarray}
\mu_B = \frac{1}{B} \sum_{b=1}^B \frac{1}{1+p_b} \ ,
\label{eq3.3}
\end{eqnarray}
while it simply reads $\mu_0 = 1$ for the eigenvalues which are unity
and which were generated by the time series not belonging to any
branch. As the weights satisfy $p_b>0$ if $b=1,\ldots,B$, we always
have
\begin{eqnarray}
\mu_B < 1 = \mu_0 \ .
\label{eq3.5}
\end{eqnarray} 
This important relation implies that the centers of the distributions
generated by the noise from the two set of degenerate eigenvalues are
always different from each other. The distribution due to the true
correlations will always be left of that which is due to trivial
self--correlation for $k=l$. In the numerical simulation, we chose
$p_b=1-1/\kappa_b$. Assuming that the sizes $\kappa_b$ are large, we
find to leading order
\begin{eqnarray}
\mu_B = \frac{1}{B} \sum_{b=1}^B \frac{1}{2-1/\kappa_b} 
      = \frac{1}{2} + \frac{1}{4B}
                  \sum_{b=1}^B \frac{1}{\kappa_b} \ ,
\label{eq3.6}
\end{eqnarray}
which explains why the distribution due to the true correlations is
roughly centered around $1/2$ in figure~\ref{fig_CEvL_T}.

\begin{figure}
\begin{indented}
\item[]
\begin{minipage}{7.5cm}
\psfig{figure=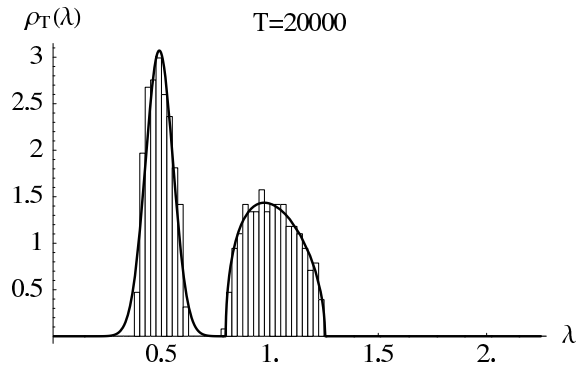,width=7.5cm,angle=0}
\end{minipage}
\end{indented}
\caption{\label{fig_CEvL3DP_T}
         Bulk part of the spectral density $\rho_{20000}(\lambda)$ for
         $T=20000$. The phenomenological formula~(\ref{eq3.11}) is
         fitted to it. The density is given in units of $K$.}
\end{figure}

To find an estimate for the noisy density $\rho_T(\lambda)$, we argue
phenomenologically. Some further analytical properties are compiled
in~\ref{appA}.  We first replace the first term of Eq.~(\ref{eq3.1})
with a Gaussian distribution centered at $\mu_B$,
\begin{eqnarray}
G(\lambda-\mu_B,v_B^2/T) = \sqrt{\frac{T}{2\pi v_B^2}}
                    \exp\left(-\frac{(\lambda-\mu_B)^2}
                                    {2v_B^2/T}\right) \ .
\label{eq3.7}
\end{eqnarray}
For its variance, we write $v_B^2/T$ where $v_B$ is a proper geometric
average of those numbers $v_k$ in Eq.~(\ref{eq3.2}) which stem from
the industrial branches.  As a weight factor, we sum over the
multiplicities in the degenerate case~(\ref{eq3.1}),
\begin{eqnarray}
\sum_{b=1}^B (\kappa_b-1) = K - \kappa - B \ .
\label{eq3.8}
\end{eqnarray} 
Similarly, we could replace the third term of Eq.~(\ref{eq3.1}) with a
Gaussian centered at $\mu_0=1$. However, it is better to employ the
fact the $\kappa\times\kappa$ block of the companies which are in no
branch belongs to a chiral random matrix ensemble~\cite{TWettig,Gu2}
whose density is known~\cite{Dys,Seng} to have the algebraic form
\begin{eqnarray}
\rho_{\rm ch}(\lambda,\kappa/T) = \frac{T}{2\pi\kappa}
     {\rm Re\,}\frac{\sqrt{(\lambda_+-\lambda)
                     (\lambda-\lambda_-)}}{\lambda} \ .
\label{eq3.9}
\end{eqnarray} 
where $\lambda_\pm$ are the largest and the smallest eigenvalues,
respectively, given by
\begin{eqnarray}
\lambda_\pm = 1+\frac{\kappa}{T}
                           \pm 2\sqrt{\frac{\kappa}{T}} \ .
\label{eq3.10}
\end{eqnarray}
Both of them converge to $\mu_0=1$ for $T\to\infty$.  Outside the
interval $\lambda_-<\lambda<\lambda_+$, the square root in
Eq.~(\ref{eq3.9}) is strictly imaginary. Thus, the real part ensures
that the function $\rho_{\rm ch}(\lambda,\kappa/T)$ is zero outside
the supporting interval $\lambda_-<\lambda<\lambda_+$.  As a weight
factor, we take the multiplicity $\kappa$ in the degenerate
case~(\ref{eq3.1}). We notice that, analogous to $v_B$, there would be
in principle a scale $v_0$ entering the density $\rho_{\rm
ch}(\lambda,\kappa/T)$. It would result from a proper average of those
numbers $v_k$ in Eq.~(\ref{eq3.2}) which do not stem from the
industrial branches. However, since we start out from normalized time
series and since the weights $p_b$ do not contribute here, we expect
$v_0=1$.  The second term of Eq.~(\ref{eq3.1}) we leave unchanged,
because we can ignore the shift in the positions of these largest
eigenvalues. Thus, collecting everything, we find
\begin{eqnarray}
\rho_T(\lambda)
 &=& (K-\kappa-B)G(\lambda-\mu_B,v_B^2/T) 
                         \nonumber\\
 & & 
   + \sum_{b=1}^B \delta\left(\lambda-\frac{1+\kappa_bp_b}{1+p_b}\right) 
   + \kappa \rho_{\rm ch}(\lambda,\kappa/T) 
                                                  \ .
\label{eq3.11}
\end{eqnarray}
As the functions $G(\lambda-\mu_B,v_B^2/T)$ and $\rho_{\rm
ch}(\lambda,\kappa/T)$ are normalized to unity, the noisy density
$\rho_T(\lambda)$ is correctly normalized to the total number $K$ of
eigenvalues. We expect the formula~(\ref{eq3.11}) to work particularly
well for large $T$. In figure~\ref{fig_CEvL3DP_T} we fit it to the bulk
part of the spectral density $\rho_{20000}(\lambda)$ for $T=20000$
which was presented in figure~\ref{fig_CEvL_T}. The agreement is very
good, the fit yields $v_B=9$.

\section{Noise Identification by Power Mapping}
\label{sec4}

In Sec.~\ref{sec4.1}, we introduce the power mapping and discuss its
properties numerically. We give a qualitative analytical explanation
in Sec.~\ref{sec4.2}. In Sec.~\ref{sec4.3}, we demonstrate how the power
mapping can detect different correlation structures. Finally, we
define a measure for the noise in Sec.~\ref{sec4.4}.

\subsection{Power Mapping}
\label{sec4.1}

The true correlations buried under the noise become visible in the
spectral density if the time series are long enough. Thus, if we found
a procedure that is equivalent, in some sense, to a prolongation of
the time series, we would be able to identify and quantify the noise
in a given correlation matrix. In the sequel, we develop such a
procedure, the power mapping. We map the correlation matrix $C(T)$
to the matrix $C^{(q)}(T)$. Here, $q$ is a positive number and the
elements of $C^{(q)}(T)$ are calculated according to the definition
\begin{eqnarray}
C_{kl}^{(q)}(T) = {\rm sign\,}(C_{kl}(T)) |C_{kl}(T)|^q \ .
\label{eq4.1}
\end{eqnarray}
Thus, the mapping preserves the sign of the matrix element $C_{kl}(T)$
and raises the modulus of it to the $q$--th power.  We notice that
$C^{(q)}(T)$ is the matrix of the powers of the elements of $C(T)$,
but not the power of the matrix $C(T)$. This is crucial, because the
spectra of $C(T)$ and $(C(T))^q$ are, for $q\ne 0$, related in a
simple way: if $\lambda_k, \ k=1,\ldots,K$ are the eigenvalues of
$C(T)$, $\lambda_k^q, \ k=1,\ldots,K$ are the eigenvalues of
$(C(T))^q$. The spectrum of $C^{(q)}(T)$ is much more complicated and
depends on the eigenvalues and the eigenvectors of $C(T)$.  However,
depending on the numerical value chosen for $q$, it allows one to
suppress, in $C^{(q)}(T)$, certain elements of $C(T)$. This will now
be demonstrated through a numerical study of the spectral densities.

\begin{figure}
\begin{minipage}{7.5cm}
\psfig{figure=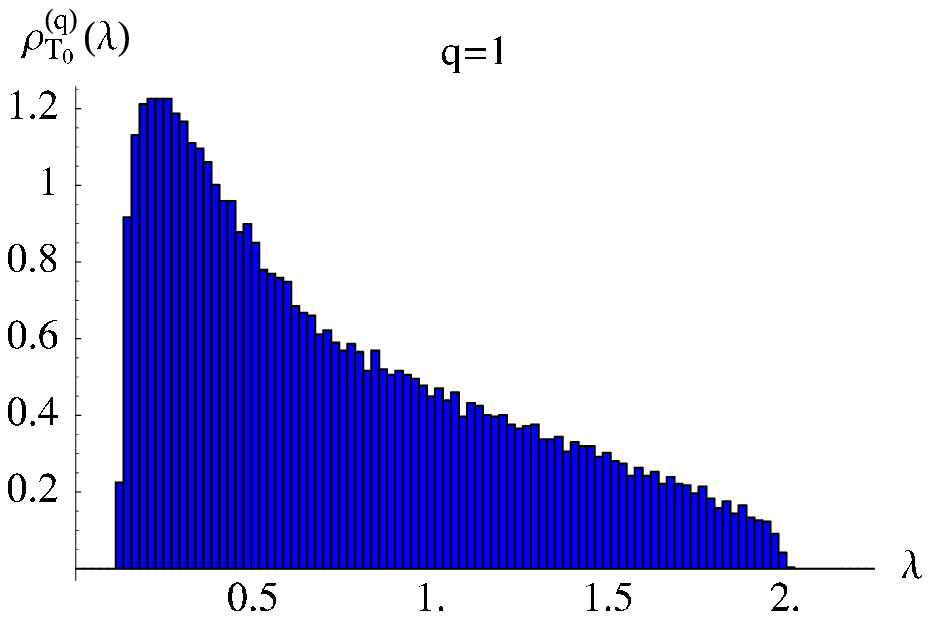,width=7.5cm,angle=0}
\end{minipage}
\hspace{0.4cm}
\begin{minipage}{7.5cm}
\psfig{figure=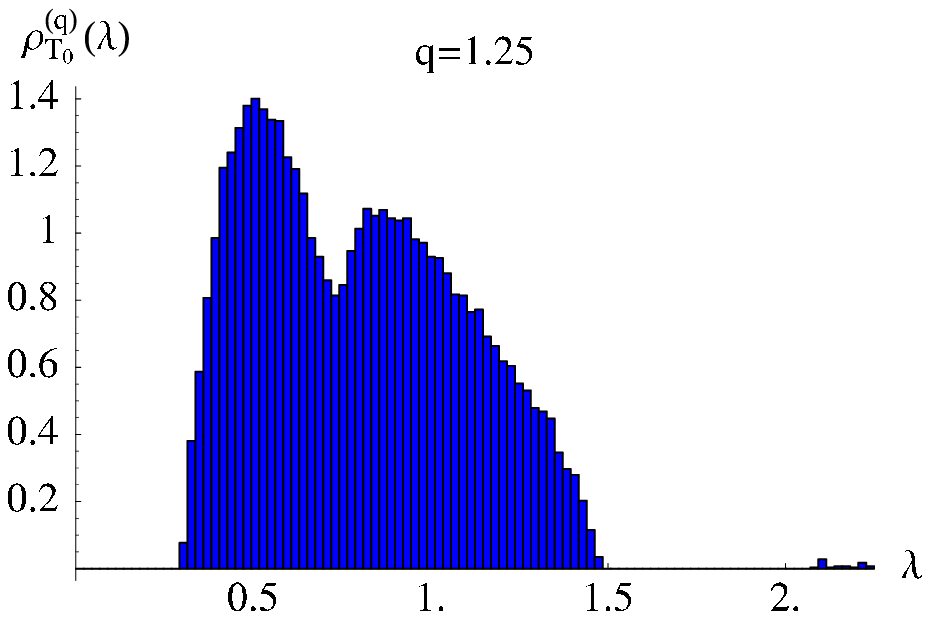,width=7.5cm,angle=0}
\end{minipage}\\ 
\vspace{0.4cm}\\
\begin{minipage}{7.5cm}
\psfig{figure=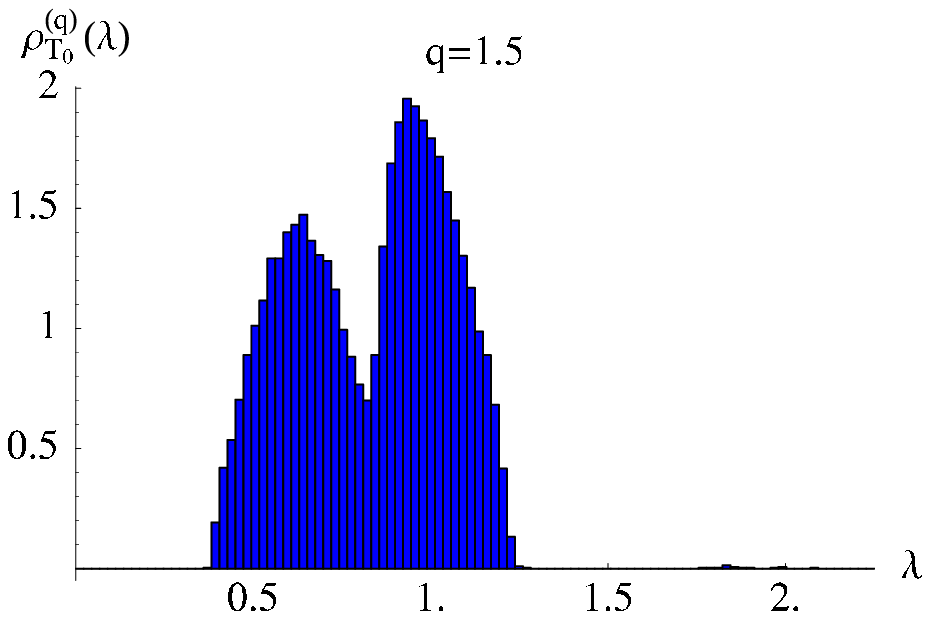,width=7.5cm,angle=0}
\end{minipage}
\hspace{0.4cm}
\begin{minipage}{7.5cm}
\psfig{figure=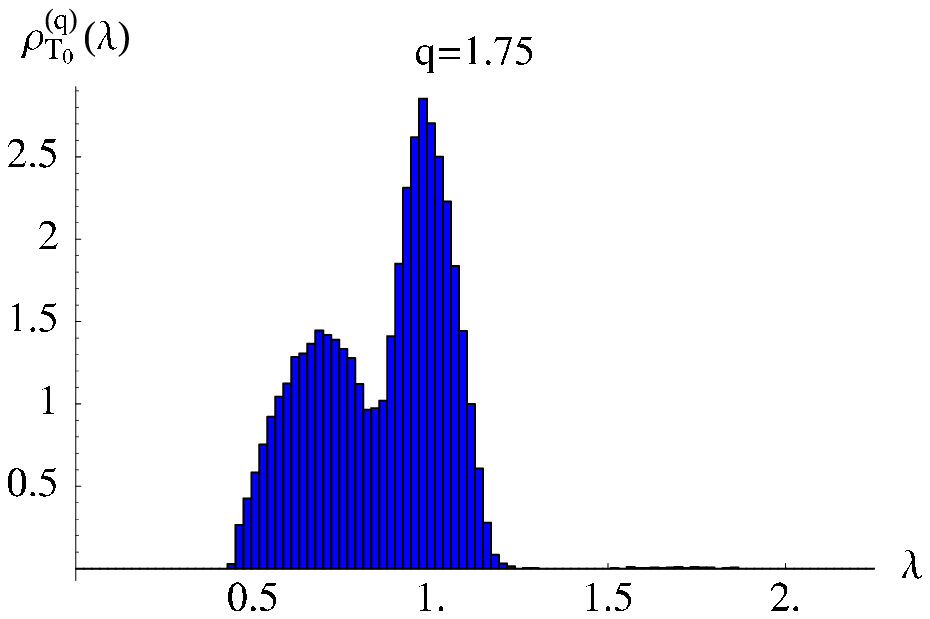,width=7.5cm,angle=0}
\end{minipage}\\ 
\vspace{0.4cm}\\
\begin{minipage}{7.5cm}
\psfig{figure=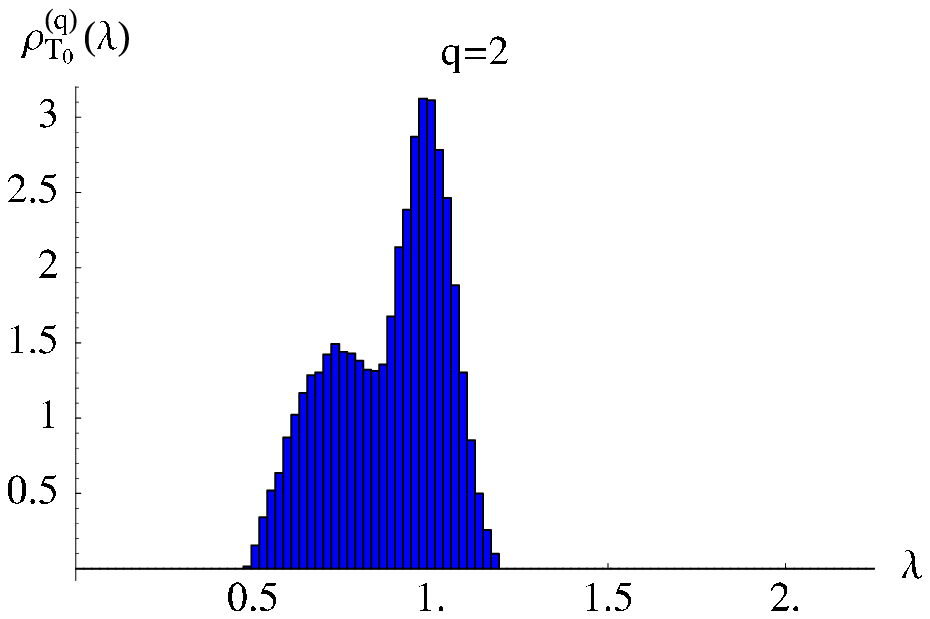,width=7.5cm,angle=0}
\end{minipage}
\hspace{0.4cm}
\begin{minipage}{7.5cm}
\psfig{figure=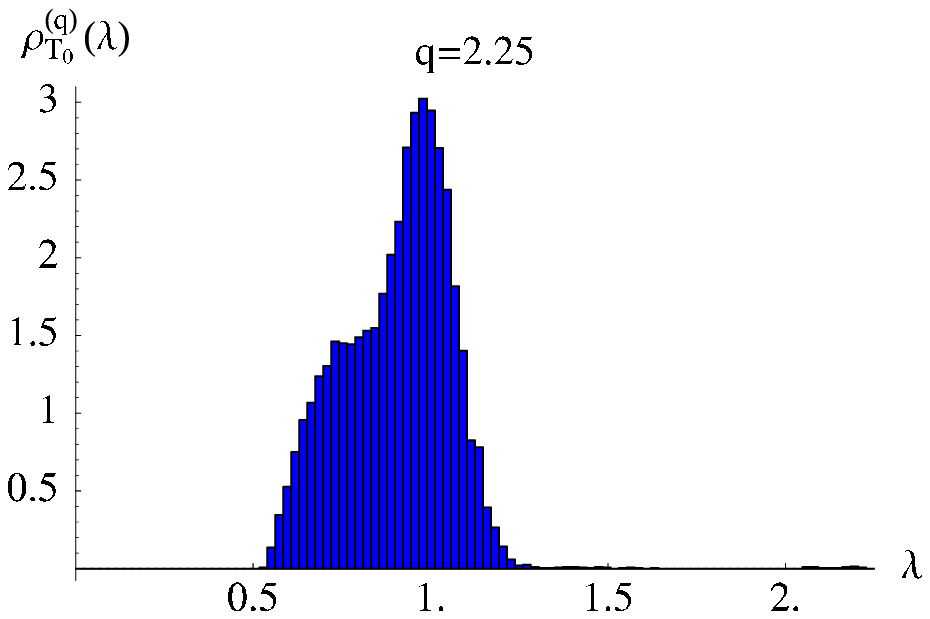,width=7.5cm,angle=0}
\end{minipage}\\ 
\vspace{0.4cm}\\
\begin{minipage}{7.5cm}
\psfig{figure=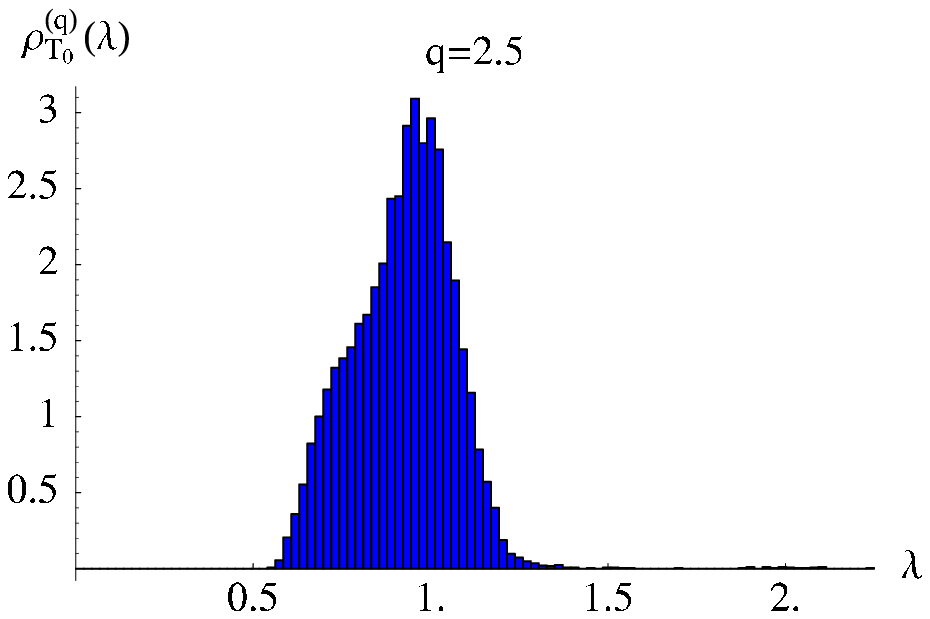,width=7.5cm,angle=0}
\end{minipage} 
\hspace{0.4cm}
\begin{minipage}{7.5cm}
\end{minipage}\\
\caption{\label{fig_EvANP_T} 
Spectral densities $\rho_{T_0}^{(q)}(\lambda)$ of the power mapped
correlation matrices $C^{(q)}$.  The length of the time series is
always $T_0=1650$. The values of the powers used are
$q=1.0,1.25,1.5,1.75,2.0,2.25,2.5$. An ensemble of 25 matrices $C$ was
simulated which were power mapped onto $C^{(q)}$.  The densities are
given in units of $K$.}
\end{figure}

We simulate correlation matrices for $K=508$ companies from time
series of length $T_0=1650$ with the sizes $\kappa_b$ and the weights
$p_b$ given in table~\ref{tab_NohP2}. We do so to make possible a
comparison with figure~\ref{fig_CEvL_T}, where the impact of
increasing length for the time series was studied, starting from a
correlation matrix with the same sizes and weights and for the same
length $T=1650$. To improve the statistical significance and to better
illustrate the effect of the power mapping, we simulate an ensemble of
25 such correlation matrices from time series of length $T_0=1650$.
We choose powers $q=1.0,1.25,1.5,1.75,2.0,2.25,2.5$ and calculate, for
a fixed $q$, the 25 power mapped matrices $C^{(q)}(T)$ from the 25
matrices $C(T)$ of the ensemble. Then, the individual spectra are
evaluated and the density $\rho_{T_0}^{(q)}(\lambda)$ results as the
ensemble average. All densities are shown in figure~\ref{fig_EvANP_T}.

The power mapping transforms the original density
$\rho_{T_0}^{(1)}(\lambda)=\rho_{T_0}(\lambda)$ for $q=1$ into
densities $\rho_{T_0}^{(q)}(\lambda)$ which show for intermediate
values of $q$ two clearly separated peaks. The best separation is
obtained near $q=1.5$. For values beyond $q=2$, the two peaks glue
together again, and the separation is lost. What do these two peaks in
the densities $\rho_{T_0}^{(q)}(\lambda)$ for $1.25 \le q \le 2.0$
represent? --- Comparison with figure~\ref{fig_CEvL_T} shows that,
indeed, the left peak in figure~\ref{fig_EvANP_T} corresponds to the
one for the true correlations in figure~\ref{fig_CEvL_T}, while the
right peak in figure~\ref{fig_EvANP_T} corresponds to the one for the
noise around the trivial self--correlation of the companies for $k=l$
in figure~\ref{fig_CEvL_T}. Hence, we have found the desired procedure
which roughly amounts to a prolongation of the time series.

\subsection{Qualitative Analytical Discussion}
\label{sec4.2}

To leading order in the length $T$ of the time series, the $C_{kl}(T)$
are given by Eq.~(\ref{eq2.9}).  To understand the effect of the power
mapping, we distinguish three different cases in considering
$|C_{kl}(T)|^q$. First, we power map the diagonal elements
$C_{kk}(T)$. As Eq.~(\ref{eq2.9}) shows, the vast majority of them
matrix elements will be positive if $T$ is sufficiently large.
Thus, to simplify the discussion, we ignore the absolute value sign.
To leading order in $T$, we have
\begin{eqnarray}
(C_{kk}(T))^q &=& 1 + \frac{q}{1+p_{b(k)}} 
                      \Biggl(\sqrt{2}p_{b(k)}a_{b(k)b(k)}
                               + \sqrt{2}a_{kk}
                         \nonumber\\
              & & \qquad
                    + \sqrt{p_{b(k)}}(a_{b(k)k}+a_{kb(k)})\Biggr)
                                 \frac{1}{\sqrt{T}} \ .
\label{eq4.11}
\end{eqnarray}
Second, we power map the off--diagonal elements $C_{kl}(T)$ in the
blocks of the industrial branches where $k\ne l$ but $b(k)=b(l)$.  For
the same reason as in the previous case, we ignore the absolute value
sign, and find
\begin{eqnarray}
(C_{kl}(T))^q &=&  \left(\frac{p_{b(k)}}{1+p_{b(k)}}\right)^q 
   + \frac{q(p_{b(k)})^{q-1}}{(1+p_{b(k)})^q} 
                      \Biggl(\sqrt{2}p_{b(k)}a_{b(k)b(k)}
                               + \sqrt{2}a_{kl}
                         \nonumber\\
              & & \qquad
                    + \sqrt{p_{b(k)}}(a_{b(k)l}+a_{kb(l)})\Biggr)
                                 \frac{1}{\sqrt{T}} \ ,
\label{eq4.12}
\end{eqnarray}
to leading order in $T$. Third, we power map the elements $C_{kl}(T)$
outside the blocks, where $k\ne l$ and $b(k)\ne b(l)$. Since all Kronecker 
$\delta$'s in Eq.~(\ref{eq2.9}) are zero in this case, we obtain
\begin{eqnarray}
|C_{kl}(T)|^q &=& \frac{1}{\left((1+p_{b(k)})(1+p_{b(l)})\right)^{q/2}}
         \Bigg|\sqrt{p_{b(k)}}\sqrt{p_{b(l)}}a_{b(k)b(l)} + a_{kl} 
                         \nonumber\\
              & & \qquad + \sqrt{p_{b(k)}}a_{b(k)l}
                         + \sqrt{p_{b(l)}}a_{kb(l)}\Bigg|^q
                           \frac{1}{T^{q/2}}
\label{eq4.13}
\end{eqnarray}
as the leading order in $T$. We notice that the matrix elements
$C_{kl}(T)$ in this third case will be positive or negative with equal
probability.

In the first two cases~(\ref{eq4.11}) and~(\ref{eq4.12}), the powers
$(C_{kl}(T))^q$ contain a $T$ independent term plus a term which is of
order $1/\sqrt{T}$. In the third case, however, there is no $T$
independent term, and the leading order of the whole expression is
$1/T^{q/2}$. Thus, for $q>1$, $|C_{kl}(T)|^q$ vanishes faster in the
third case than in the first two terms. As the case~(\ref{eq4.13})
comprises all elements outside the blocks, where $k\ne l$ and $b(k)\ne
b(l)$, we find that, for $q>1$, the power mapped matrix $C^{(q)}(T)$
is block diagonal to leading order $1/\sqrt{T}$. This explains why the
power mapping has an effect which is comparable to a prolongation of
the time series. At first sight, one would expect that the effect is
the stronger, the larger the value of $q$.  However, this is not so,
because the $T$ independent terms are different for the matrix
elements in the blocks: in the first case~(\ref{eq4.11}) of the
diagonal elements, it is unity, but a number smaller than unity in the
second case~(\ref{eq4.12}). The larger the value of $q$, the smaller
becomes the latter term.  Hence, the unity in the diagonal elements
dominates more and more, driving the eigenvalues towards unity.  The
effect discussed above which is comparable to a prolongation of the
time series, leads to the separation of the spectral densities into two
peaks. If $q$ becomes too large, this is counteracted, and the
two peaks merge again. Consequently, there must be an optimal
values for $q$. In the numerical simulation we found that it is
roughly $q=1.5$.

For infinitely long time series $T\to\infty$, the power mapped
correlation matrix $C^{(q)}(\infty)$ is trivially block diagonal.  The
eigenvalues $\lambda_k^{(q)}(\infty)$ and the spectral density
$\rho_\infty^{(q)}(\lambda)$ are found along lines similar to the ones
leading to Eq.~(\ref{eq3.1}). We arrive at
\begin{eqnarray}
\rho_\infty^{(q)}(\lambda)
 &=& \sum_{b=1}^B (\kappa_b-1)\delta\left(\lambda-
        \left(1-\left(\frac{p_b}{1+p_b}\right)^q\right)\right)
                         \nonumber\\
 & & 
   + \sum_{b=1}^B \delta\left(\lambda-
  \left(1+(\kappa_b-1)\left(\frac{p_b}{1+p_b}\right)^q\right)\right)
   + \kappa \delta\left(\lambda-1\right) \ .
\label{eq4.14}
\end{eqnarray} 
This correctly reduces to Eq.~(\ref{eq3.1}) for $q=1$. As already
argued, the peaks due to the first term move towards the peak due to
the third term if $q$ is made large. Again, the $\delta$ peaks are
smeared out if the length of the time series is finite. We show
in~\ref{appB} that the noisy spectral density $\rho_T^{(q)}(\lambda)$
of the power mapped correlation matrix can be estimated by
\begin{eqnarray}
\rho_T^{(q)}(\lambda)
 &=& (K-\kappa-B)G(\lambda-\mu_B^{(q)},(v_B^{(q)})^2/T) 
                         \nonumber\\
 & & 
  + \sum_{b=1}^B \delta\left(\lambda-
  \left(1+(\kappa_b-1)\left(\frac{p_b}{1+p_b}\right)^q\right)\right)
                         \nonumber\\
 & & 
   + \kappa G(\lambda-1,(v_0^{(q)})^2/T^q) \ .
\label{eq4.15}
\end{eqnarray}
Here, the parameters $(v_B^{(q)})^2$ and $(v_0^{(q)})^2$ collect the
effects due to the smearing out and
\begin{eqnarray}
\mu_B^{(q)} = \frac{1}{B} \sum_{b=1}^B \left(1-
                    \left(\frac{p_b}{1+p_b}\right)^q\right)
            = 1-\frac{1}{B} 
                \sum_{b=1}^B \left(\frac{p_b}{1+p_b}\right)^q 
\label{eq4.16}
\end{eqnarray}
approximates the center of the Gaussian due the true correlations in
the first term of Eq.~(\ref{eq4.15}).  The third term estimates the
peaks due to the noise.  As discussed in~\ref{appB}, these two
contributions are affected differently as finite lengths of the time
series are concerned. Thus, it is useful to scale the variance with
$1/T$ in the first term, but with $1/T^q$ in the third one.

\subsection{Detecting Different Correlation Structures}
\label{sec4.3}

\begin{figure}
\begin{indented}
\item[]
\begin{minipage}{6cm}
\psfig{figure=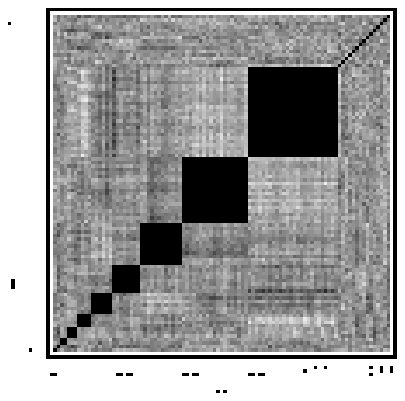,width=6cm,angle=0}
\end{minipage}
\end{indented}
\vspace{0.7cm}
\begin{indented}
\item[]
\begin{minipage}{6cm}
\psfig{figure=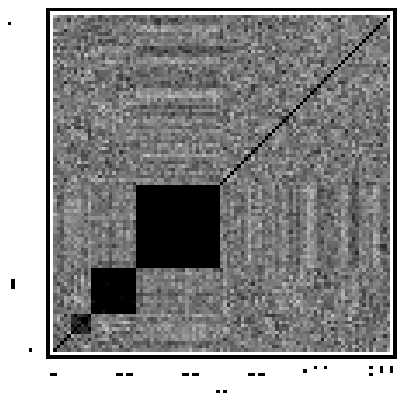,width=6cm,angle=0}
\end{minipage}
\end{indented}
\vspace{0.7cm}
\begin{indented}
\item[]
\begin{minipage}{6cm}
\psfig{figure=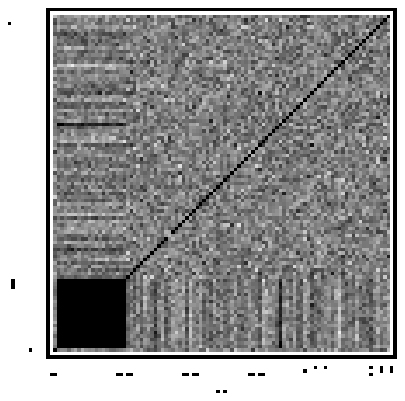,width=6cm,angle=0}
\end{minipage}
\end{indented}
\vspace{0.7cm}
\caption{\label{fig_CMP_T}
Three correlation matrices $C$ with different correlation
structure.  The parameters for the top, the middle and the bottom
structure, respectively, are given in table~\protect{\ref{tab_CM}}. The
amount of noise in the middle and the bottom structure is much higher
than in the top structure.}
\end{figure}

\begin{figure}
\begin{minipage}{7.5cm}
\psfig{figure=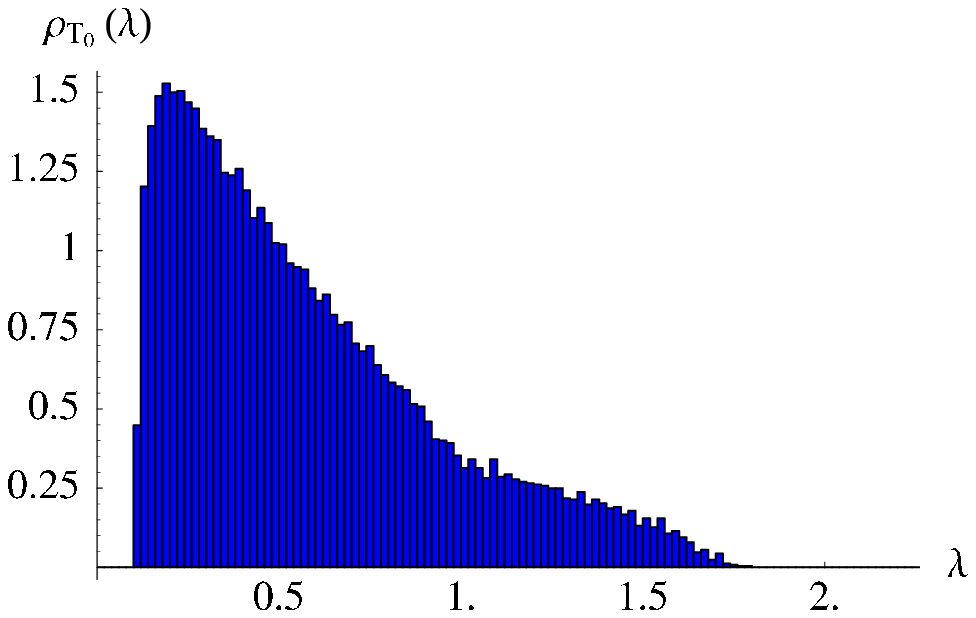,width=7.5cm,angle=0}
\end{minipage}
\hspace{0.2cm}
\begin{minipage}{7.5cm}
\psfig{figure=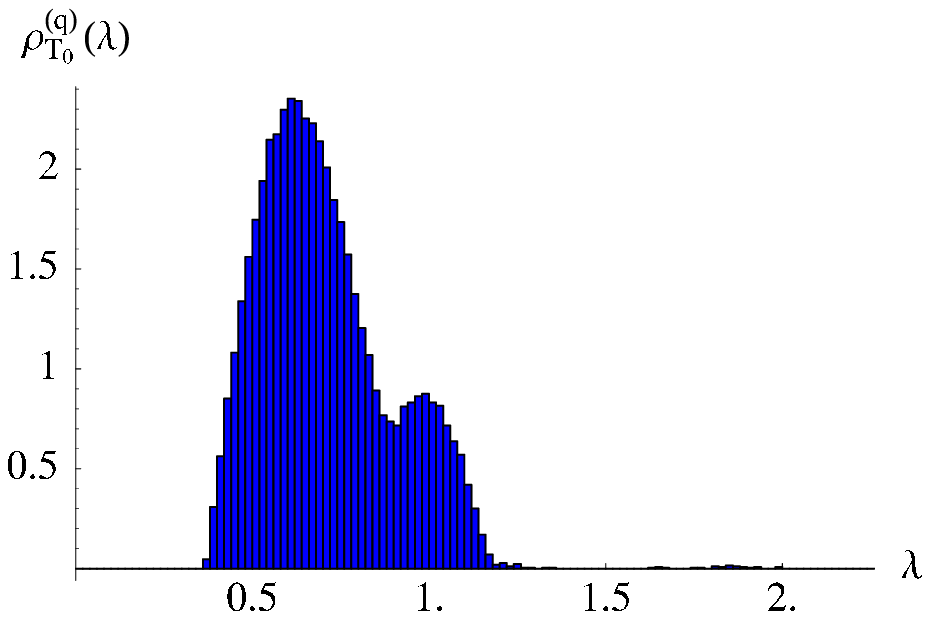,width=7.5cm,angle=0}
\end{minipage}\\ 
\vspace{0.4cm}\\
\begin{minipage}{7.5cm}
\psfig{figure=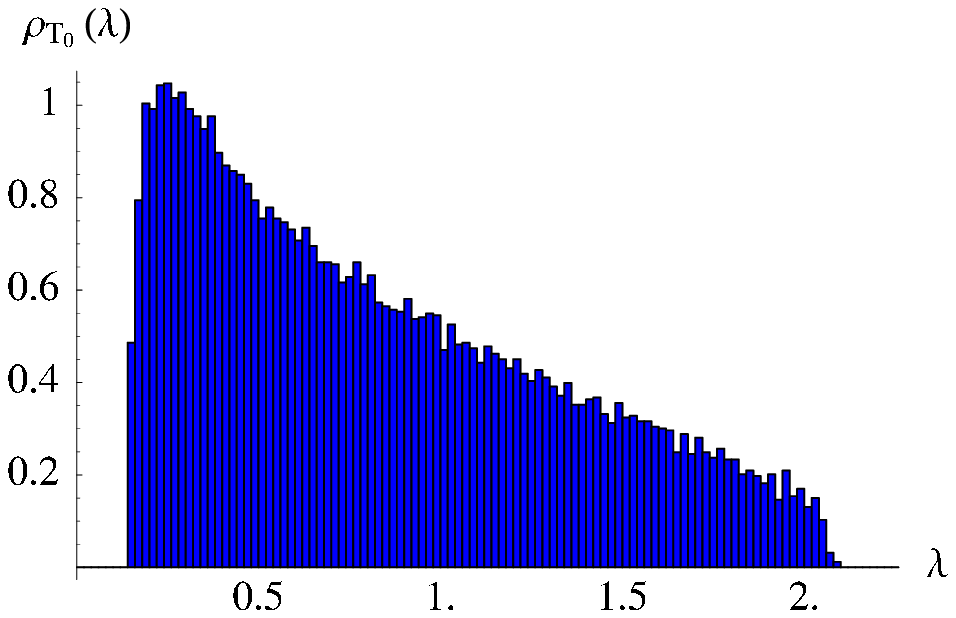,width=7.5cm,angle=0}
\end{minipage}
\hspace{0.2cm}
\begin{minipage}{7.5cm}
\psfig{figure=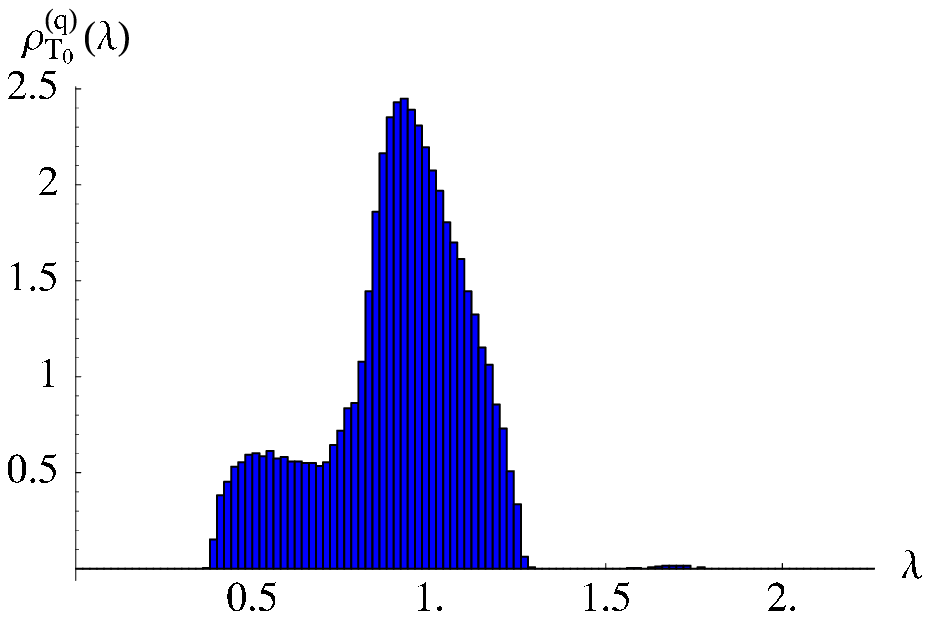,width=7.5cm,angle=0}
\end{minipage}\\ 
\vspace{0.4cm}\\
\begin{minipage}{7.5cm}
\psfig{figure=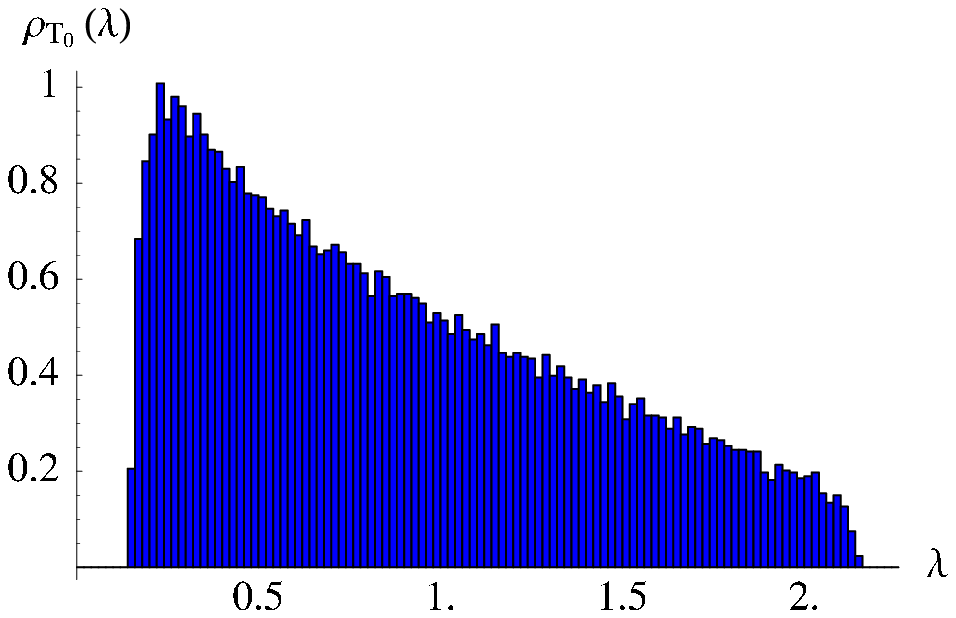,width=7.5cm,angle=0}
\end{minipage}
\hspace{0.2cm}
\begin{minipage}{7.5cm}
\psfig{figure=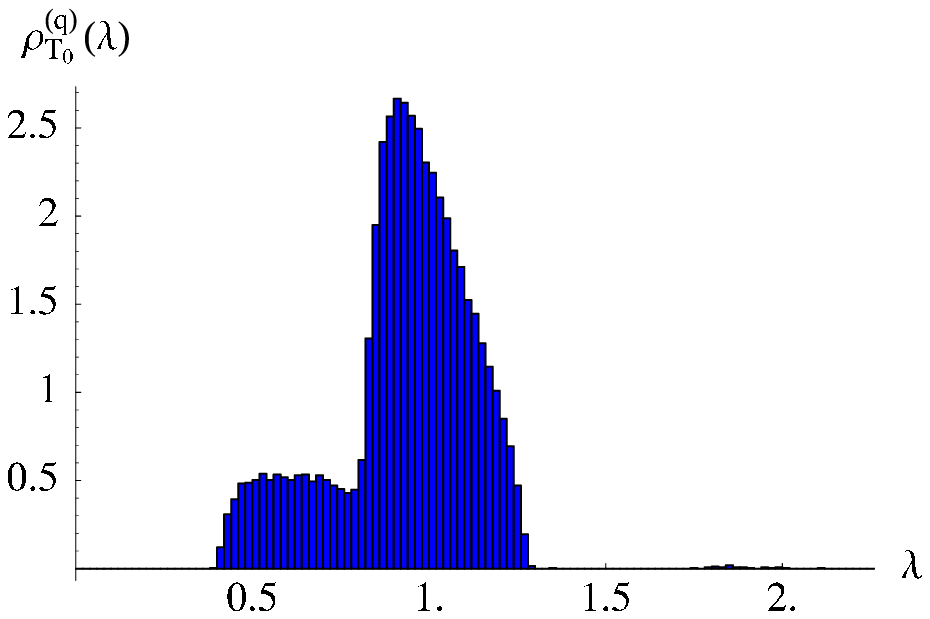,width=7.5cm,angle=0}
\end{minipage}\\
\caption{\label{fig_EvEP_T} 
Left column: Spectral densities $\rho_{T_0}(\lambda)$ of the original
correlation matrices $C$.  Right column: Spectral densities
$\rho_{T_0}^{(q)}(\lambda)$ of the same, but power mapped correlation
matrices $C^{(q)}$.  The length of the time series is always
$T_0=1650$.  An ensemble of 25 matrices $C$ was simulated for each
structure.  The value of the power used is always $q=1.5$. The 25
matrices $C$ which were simulated for each structure were individually
power mapped onto matrices $C^{(q)}$, forming a new ensemble for each
structure. The densities are given in units of $K$.}
\end{figure}

We consider three examples for different correlation structures.  The
parameters are listed in table~\ref{tab_CM}, and the corresponding
correlation matrices $C$ are displayed in figure~\ref{fig_CMP_T}. We
refer to the three examples as top, middle and bottom structure,
respectively. 
\begin{table}
\caption{\label{tab_CM}
         The sizes $\kappa_b$ and the weights $p_b$ for the industrial
         branches used in the numerical simulation for the three
         ensembles of correlation matrices with different structures.
         The total dimension of the matrices is always $K=508$.}
\vspace{0.3cm}
\begin{indented}
\item[]{\small Top structure: the number of industrial branches is
$B=11$, $\kappa=76$ companies are in no branch.}
\end{indented}
\begin{indented}
\lineup
\item[]\begin{tabular}{c|ccccccccccc}
\br 
$b$          & 1 & 2 &  3 &  4 &  5 & 6 & 7 & 8 & 9 & 10 & 11 \\ 
\mr
$\kappa_{b}$ & 2 & 4 & 7 & 10 & 15 & 20 & 30 & 42 & 64 & 98 & 140 \\
$p_b$        & 0.5 & 0.75 & 0.85 & 0.9 & 0.93 & 0.95 & 0.96 
             & 0.97 & 0.984 & 0.989 & 0.99 \\ 
\br
\end{tabular}
\end{indented}
\vspace{0.3cm}
\begin{indented}
\item[]{\small Middle structure: the number of industrial branches is
$B=6$, $\kappa=256$ companies are in no branch.}
\end{indented}
\begin{minipage}{5.0cm}
\begin{indented}
\lineup
\item[]\begin{tabular}{c|cccccc}
\br 
$b$          & 1 & 2 &  3 &  4 &  5 & 6 \\ 
\mr
$\kappa_{b}$ & 4 & 8 & 16 & 32 & 64 & 128 \\
$p_b$        & 0.008 & 0.01 & 0.03 & 0.07 & 0.2 & 0.99 \\ 
\br
\end{tabular}
\end{indented}
\end{minipage}
\begin{minipage}{3.6cm}
\end{minipage}
\vspace{0.3cm}
\begin{indented}
\item[]{\small Bottom structure: the number of industrial branches is
$B=2$, $\kappa=400$ companies are in no branch.}
\end{indented}
\begin{minipage}{2.5cm}
\begin{indented}
\lineup
\item[]\begin{tabular}{c|cc}
\br 
$b$          & 1 & 2 \\ 
\mr
$\kappa_{b}$ & 4 & 104 \\
$p_b$        & 0.75 & 0.99 \\ 
\br
\end{tabular}
\end{indented}
\end{minipage}
\begin{minipage}{6.1cm}
\end{minipage}
\end{table}
For all three structures, the total number of companies is $K=508$ and
the length of the time series is $T_0=1650$.  The top structure has
many branches ($B=11$) with relatively strong correlations and little
noise, the middle structure has less branches ($B=6$) with weaker
correlations and more noise and, finally, the bottom structure has
only $B=2$ branches with stronger correlations, and a considerable
amount of noise. We notice that the weights $p_b$ are not chosen
according to Eq.~(\ref{eq3.0}). Rather, we adjusted them in such a way
that the spectral densities $\rho_{T_0}(\lambda)$ for the middle and
the bottom structure look as similar as possible. This can be seen in
the left column of figure~\ref{fig_EvEP_T}.  The top structure serves
as a reference example. As the figure shows, the true correlations are
so dominating that the spectral density is already almost separated in
two substructures which are separated by a kink around
$\lambda=1$. Thus, the top structure is an idealizing example.

We now apply the power mapping and calculate matrices $C^{(q)}$
according to Eq.~(\ref{eq4.1}). We choose the value $q=1.5$ which we
identified as optimal in Sec.~\ref{sec4.1}. The resulting spectral
densities $\rho_{T_0}^{(q)}(\lambda)$ are shown in the right column of
figure~\ref{fig_EvEP_T}. For each of the three structures, we see the
separation into two peaks, the left one corresponding to the true
correlations, the right one to the noise. As expected, the spectral
density for the top structure consisting of a big left peak and a
small right peak differs considerably from the ones for the middle and
the bottom structures where the left peaks are small and the right
peaks big. This nicely confirms our expectation that the power mapping
is capable of efficiently distinguishing the gross structures in the
correlation matrices.

It is now interesting to compare the spectral densities for the middle
and the bottom structure: although the shape of densities
$\rho_{T_0}(\lambda)$ in the left column of figure~\ref{fig_EvEP_T}
can hardly be distinguished, the densities $\rho_{T_0}^{(q)}(\lambda)$
for the power mapped correlation matrices, shown in the right column,
have similar, but distinguishable shapes.  For both structures, the
left peak is small, the right one big. However, for the bottom
structure, the right peak is narrower and its left shoulder is
steeper. Hence, the power mapping also gives useful information about
the fine structure in the correlation matrices.

\subsection{Measuring Correlations and Noise}
\label{sec4.4}

\begin{figure}
\begin{indented}
\item[]
\begin{minipage}{7.5cm}
\psfig{figure=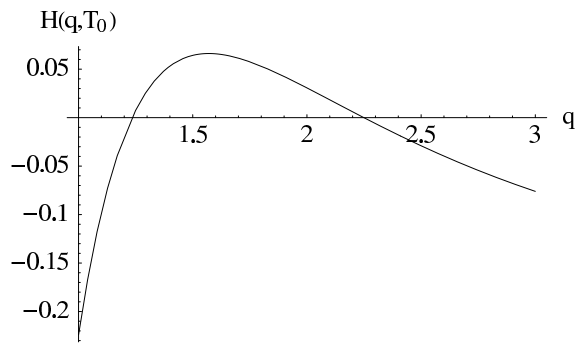,width=7.5cm,angle=0}
\end{minipage}
\end{indented}
\caption{\label{fig_hPl_T}
A typical example for the function $H(q,T)$ versus the power
$q$. The parameters for the original correlation matrix in
Sec.~\ref{sec3.1} are used with $T=T_0=1650$. The optimal $q$ value is
defined by the maximum value of $H(q,T_0)$ which is seen to be reached
near $q=1.5$.}
\end{figure}

Using the estimate~(\ref{eq4.15}), we can obtain some further
understanding of why $q=1.5$ is a good value for the power
mapping. The two Gaussian peaks which emerge from the power mapping, 
i.e.~the one due to the true correlations and the one due
to the noise, are separated if there is some space between the left
side of the former and the right side of the latter. According
to Eq.~(\ref{eq4.15}), the peaks should be separated if we have
\begin{eqnarray}
1-\frac{v_0^{(q)}}{T^{q/2}} > 
   \mu_B^{(q)} + \frac{v_B^{(q)}}{\sqrt{T}} \ ,
\label{eq4.17}
\end{eqnarray}
or, equivalently, if the function
\begin{eqnarray}
H(q,T) = \frac{1}{B} 
                \sum_{b=1}^B \left(\frac{p_b}{1+p_b}\right)^q
           -\frac{v_0^{(q)}}{T^{q/2}} 
           -\frac{v_B^{(q)}}{\sqrt{T}} 
\label{eq4.18}
\end{eqnarray}
is positive. Here, we used Eq.~(\ref{eq4.16}).  In
figure~\ref{fig_hPl_T}, we display the function $H(q,T)$ for different
power mapped correlation matrices starting from the original one used
in Sec.~\ref{sec3.1} with $T=T_0=1650$. Obviously, we may define the
optimal $q$ value as the point where $H(q,T_0)$ reaches its maximum
which is given by the equation $\partial H(q,T_0)/\partial
q=0$. Indeed, this value is close to $q=1.5$.  The parameters
$v_0^{(q)}$ and $v_B^{(q)}$ used in figure~\ref{fig_hPl_T} were obtained
from the fitting procedure to be described in the following. As the
dependence of $H(q,T)$ on $q$ is very complicated, we do not go into a
further analytical discussion.

\begin{figure}
\begin{indented}
\item[]
\begin{minipage}{7.5cm}
\psfig{figure=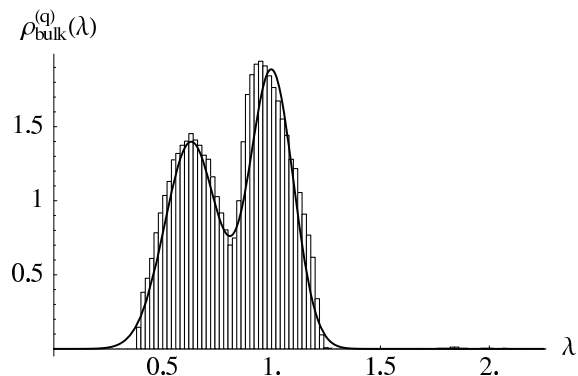,width=7.5cm,angle=0}
\end{minipage}
\end{indented}
\caption{\label{fig_RMPl12_T} 
Fit of the ansatz~(\ref{eq4.21}) to the bulk part of the spectral
density for the power mapped correlation matrix used in
Sec.~\ref{sec4.1} with $q=1.5$. The density is given in units of $K$.}
\end{figure}

To fit the spectral density of the power mapped correlation matrices,
we use an ansatz motivated by the estimate~(\ref{eq4.15}). As we only
want to fit the bulk part, we ignore the second term in the
estimate~(\ref{eq4.15}) due to the large eigenvalues and take
\begin{eqnarray}
\rho_{\rm bulk}^{(q)}(\lambda)
 = R_B^{(q)} G(\lambda-\mu_B^{(q)},(\sigma_B^{(q)})^2) 
   + R_0^{(q)} G(\lambda-1,(\sigma_0^{(q)})^2) \ .
\label{eq4.21}
\end{eqnarray}
The prefactors $R_B^{(q)}$, $R_0^{(q)}$, the standard deviations
$\sigma_B^{(q)}$, $\sigma_0^{(q)}$ and the mean value $\mu_B^{(q)}$
are fit parameters.  This relatively high number of fit parameters is
acceptable, because the data to be fitted have a clear structure.
Moreover, it should be emphasized that we are only interested in
obtaining proper estimates by these fits. From the previous
discussion, we expect that the fit should approximately give, first,
$R_B^{(q)}=K-\kappa-B$ and $R_0^{(q)}=\kappa$ for the prefactors and,
second, the scaling behavior $(\sigma_B^{(q)})^2=(v_B^{(q)})^2/T$ and
$(\sigma_0^{(q)})^2=(v_0^{(q)})^2/T^q$ for the variances.  We employ
the parameters $\sigma_B^{(q)}$ and $\sigma_0^{(q)}$ instead of the
parameters $v_B^{(q)}$ and $v_0^{(q)}$, because, in practice, one will
mostly have to deal with correlation matrices for one fixed length $T$
of the time series. As an example, we fit the ansatz~(\ref{eq4.21}) to
the spectral density of the power mapped correlation matrix discussed
in Sec.~\ref{sec4.1} for $q=1.5$. The result is shown in
figure~\ref{fig_RMPl12_T}.
We obtain for the prefactors $R_B^{(q)}=0.42K=213$,
$R_0^{(q)}=0.47K=239$, for the standard deviations
$\sigma_B^{(q)}=0.12$, $\sigma_0^{(q)}=0.10$ and for the mean value
$\mu_B^{(q)}=0.63$. Thus, we obtain the estimate
$\kappa=R_0^{(q)}=239$ for the number of companies which are not in
any branch. This is in fair agreement with the true value
$\kappa=256$. The sum $R_B^{(q)}+R_0^{(q)}=0.89K=452$ deviates by
about ten percent from the theoretical value
$R_B^{(q)}+R_0^{(q)}=K-B=502$. This is not so surprising, considering
the fact that our ansatz~(\ref{eq4.21}) can only be a rough
approximation. Hence, we suggest to use the ratio
\begin{eqnarray}
r^{(q)} = \frac{R_0^{(q)}}{R_B^{(q)}}
\label{eq4.22}
\end{eqnarray}
to characterize a correlation matrix. The larger $r^{(q)}$, the more
noise is present, the smaller $r^{(q)}$, the stronger are the true
correlations. In our example, we have $r^{(q)}=1.12$, implying that
noise and true correlations are more or less equally strong.
Advantageously, the ratio~(\ref{eq4.22}) is not so sensitive to the
total number $K$ of companies, implying that correlation matrices of
different sizes $K$ can have the same $r^{(q)}$. Of course, the number
$K$, the dimension of the correlation matrix, is always known in an
analysis. In the previous Sec.~\ref{sec4.3}, we discussed a
correlation matrix, labeled ``middle structure'', also involving
$\kappa=256$ companies which are in no branch. Most of the
correlations within the branches are so weak that the left peak in the
middle of the right column in figure~\ref{fig_EvEP_T} is considerably
smaller than the one in figure~\ref{fig_RMPl12_T}. The ratio $r^{(q)}$
makes the desired distinction: the overall strength of the true
correlations is much weaker than in the present case. In addition, the
standard deviation $\sigma_B^{(q)}$ and the mean value $\mu_B^{(q)}$
yield information about the spreading of the weights $p_b$ and about
their average.

Finally, we test the quality of the scaling behavior
$\sigma_B^{(q)}\propto 1/\sqrt{T}$ and $\sigma_0^{(q)}\propto
1/T^{q/2}$ for the standard deviations.  To this end, we generate
correlation matrices for various lengths $T$ of the time series and
power map them using $q=1.0$, $q=1.25$ and $q=1.5$. We fit the
resulting spectral densities using the ansatz~(\ref{eq4.21}), extract
the standard deviations $\sigma_B^{(q)}$ and $\sigma_0^{(q)}$, and fit
the latter to the expected scaling behavior $1/\sqrt{T}$ and
$1/T^{q/2}$, respectively. The expectation is well confirmed for
$\sigma_B^{(q)}$. In the case of $\sigma_0^{(q)}$, the general trend
is reproduced for the three $q$ values. For $q=1.5$, however, the most
interesting value, the agreement is good.  This is encouraging,
because the steps that led to the estimate~(\ref{eq4.15}) involved
various approximations. In any case, as already argued, the practical
applicability of our ansatz~(\ref{eq4.21}) is not affected by these
scaling questions.
\begin{figure}
\begin{minipage}{7.5cm}
\psfig{figure=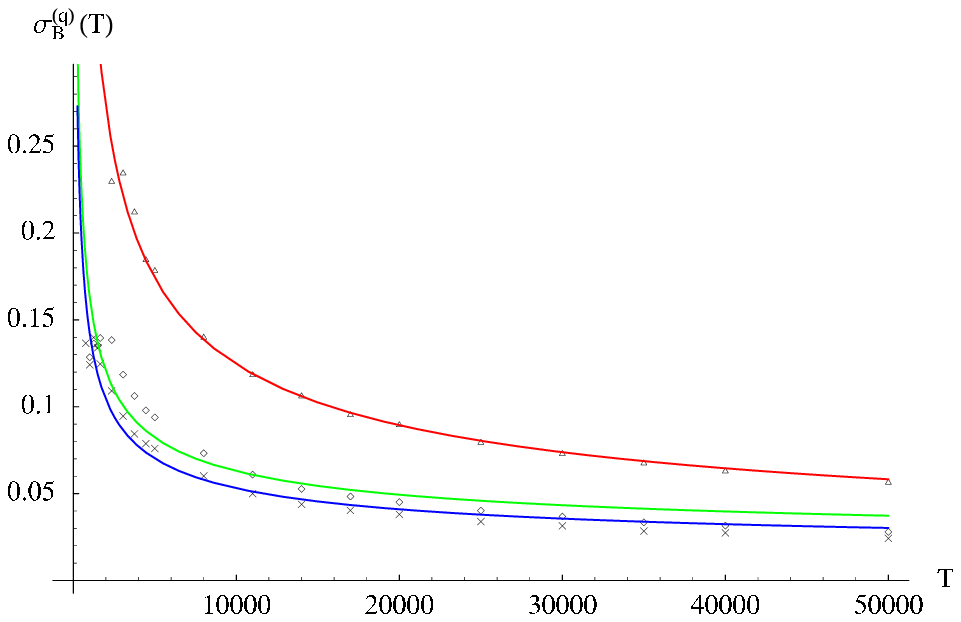,width=7.5cm,angle=0}
\end{minipage}
\hspace{0.4cm}
\begin{minipage}{7.5cm}
\psfig{figure=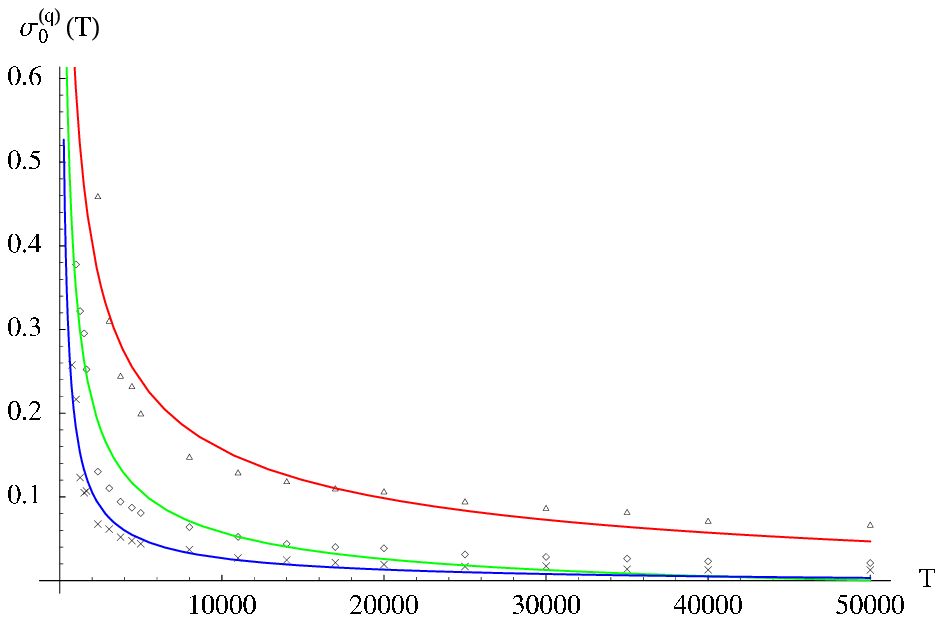,width=7.5cm,angle=0}
\end{minipage}\\
\caption{\label{fig_TP1_T}
Standard deviations $\sigma_B^{(q)}$ (top part) and
$\sigma_0^{(q)}$ (bottom part) versus the length $T$ of the time
series. The dots are the data points obtained from the fits to the
spectral densities. The solid lines are the fits of the expected
scaling behavior $1/\sqrt{T}$ and $1/T^{q/2}$ to these data
points. The top, middle and bottom curves correspond to $q=1.0$,
$q=1.25$ and $q=1.5$, respectively.}
\end{figure}

As figure~\ref{fig_TP1_T} shows, both standard deviations
$\sigma_B^{(q)}$ and $\sigma_0^{(q)}$ become smaller as $q$ is made
larger while $T$ is kept fixed. This is why the power mapping can be
viewed as effectively prolonging the time series: To obtain given
values for the standard deviations, one can either keep $q=1$ and make
the time series longer or make $q$ larger and leave $T$ unchanged. For
example, compared to the correlation matrix with $T=1000$ and $q=1$,
the noise is reduced by 60\% when one goes to $T=10000$ without
changing $q=1$ and it is reduced by 75\% when one applies the power
mapping with $q=1.5$ while keeping $T=1000$. This is most important
for applications, because time series for stocks with $T=10000$, say,
are seldom available, while time series with $T=1000$ are easily
accessible. Thus, power mapping can lead to better estimates for the
risk and for related quantities such as the
Value--at--Risk~\cite{Deutsch,RM}.

\section{Summary and Conclusion}
\label{sec5}

We developed a new method to identify and estimate the noise and,
hence, also the strength of the true correlations in a financial
correlation matrix.  The essence of our approach is the power
mapping. It suppresses those matrix elements which can be associated
with the noise. Our approach could be seen as an generalization and
extension of shrinking techniques~\cite{RD} used in contexts different
from the present one.  Importantly, the spectral density changes
drastically due to the suppression. The method itself yields a
criterion to choose the optimal power. Loosely speaking, our method
can be interpreted as an effective prolongation of the time
series. This feature makes it particularly suited for problems in
which, for whatever reason, only relatively short time series are
available.  We expect that our approach, which is new to the best of
our knowledge, can also be useful in other time series
problems. Although we presented our method in the framework of stocks,
we emphasize that it is applicable to every correlation matrix of time
series, independent of what these time series describe: in the context
of finance, these can be stocks, interest rates, exchange rates or
other risk factors, or, in other fields of science, completely
different observables.  Moreover, advantageously, our approach does
not involve further data processing or any other additional input.

Since our method does not use the large eigenvalues due to the
branches, it is a fundamentally different alternative to the technique
of Gopikrishnan et al.~\cite{Gop} which is based on such information.
Our method automatically takes into account all branches, because all
of them will contribute to the bulk of the spectral density. Thus,
even if a large eigenvalue is not large enough to lie outside the
bulk, our method will not miss the contribution of the corresponding
branch.  We also demonstrated, how the power mapping allows one to
distinguish different correlation structures.

Further work is in progress, in particular applications to empirical
correlation matrices obtained from real market data.

\section*{Acknowledgments}

We thank L.A.N. Amaral, P. Gopikrishnan, V. Plerou, B. Rosenow,
H.E. Stanley (Boston University), T. Rupp (Yale University), 
P. Neu (Dresdner Bank, Frankfurt) and R. Dahlhaus
(Universit\"at Heidelberg) for fruitful discussions. T.G. acknowledges
support from the Heisenberg Foundation and thanks H.A. Weidenm\"uller
and the Max Planck Institut f\"ur Kernphysik (Heidelberg) for
hospitality during the initial stages of the work.

\appendix

\section{Normalizations and Features of 
         Correlation Matrices}
\label{app0}

Empirical correlation matrices for the time series $S_k(t), \
k=1,\ldots,K$ of stocks, say, are often constructed in the following
way: to remove the drift in the data, one, first, takes the logarithms
and, second, normalizes the resulting time series $G_k(t), \
k=1,\ldots,K$ to zero mean and unit standard deviation,
\begin{eqnarray}
M_k(t) = \frac{G_k(t)-\langle G_k(t)\rangle_T}
              {\sqrt{\langle G_k^2(t)\rangle_T -
                     \langle G_k(t)\rangle_T^2}} \ .
\label{eq0.1}
\end{eqnarray}
By construction, we have $\langle M_k(t)\rangle_T=1$, independently of
how the averaging procedure is defined and for all $T$. We notice that
this is not the case for our model in Eq.~(\ref{eq2.2}). Using this
normalization, the correlation matrix is given by
\begin{eqnarray}
C_{kl}(T) &=& \langle M_k(t) M_l(t)\rangle_T \nonumber\\
          &=& \frac{\langle G_k(t)G_l(t)\rangle_T -
                    \langle G_k(t)\rangle_T\langle G_l(t)\rangle_T}
              {\sqrt{\langle G_k^2(t)\rangle_T -
                     \langle G_k(t)\rangle_T^2}
               \sqrt{\langle G_l^2(t)\rangle_T -
                     \langle G_l(t)\rangle_T^2}} \ .
\label{eq0.2}
\end{eqnarray}
Again, by construction, the diagonal elements are unity,
$C_{kk}(T)=1$, independent of the averaging procedure chosen and also
for all $T$. Once more, this differs from the diagonal elements of the
correlation matrix in our model, as can be seen from Eq.~(\ref{eq2.9})
for $k=l$.

We now investigate the off--diagonal matrix elements for $k\ne l$.
To leading order in the length $T$ of the time series, we assume
\begin{eqnarray}
\langle G_k(t)\rangle_T &=& g_k + \frac{f_k}{\sqrt{T}}
                                                    \nonumber\\
\langle G_k(t)G_l(t)\rangle_T  &=& 
        \beta_{kl} + \frac{\gamma_{kl}}{\sqrt{T}} \ ,
\label{eq0.3}
\end{eqnarray}
where $g_k$, $f_k$ and $\beta_{kl}$, $\gamma_{kl}$ are constants. A
straightforward calculation then yields
\begin{eqnarray}
C_{kl}(T) &=& \frac{1}
                   {\sqrt{\beta_{kk}-g_k^2}\sqrt{\beta_{ll}-g_l^2}}
       \left(c_{kl}^{(0)}+\frac{c_{kl}^{(1)}}{\sqrt{T}}\right) 
                                \nonumber\\
c_{kl}^{(0)} &=& \beta_{kl} - g_kg_l 
                                \nonumber\\
c_{kl}^{(1)} &=& \gamma_{kl}-g_kf_l-g_lf_k 
                 - \frac{1}{2}\left(\beta_{kl}-g_kg_l\right)
                                \nonumber\\
             & & \qquad \left(\frac{\gamma_{kk}-2g_kf_k}
                                   {\beta_{kk}-g_k^2}+
                  \frac{\gamma_{ll}-2g_lf_l}{\beta_{ll}-g_l^2}\right)
\label{eq0.4}
\end{eqnarray}
to leading order in $T$. We assume that the variances for infinite $T$
are non--zero, i.e.~$\beta_{kk}-g_k^2 > 0$. As required, we have
$C_{kk}(T)=1$.  For $k\ne l$, the forms of the expansions to leading
order in Eqs.~(\ref{eq2.9}) and~(\ref{eq0.4}) coincide. In the present
discussion, we have not explicitly incorporated a structure due
to branches. Such information would appear, for example, in the
parameters $\beta_{kl}$. 

As compared to the correlation matrices resulting from the model in
Eq.~(\ref{eq2.2}), the difference in the normalization of the diagonal
elements in Eq.~(\ref{eq0.2}) has only a marginal and negligible
influence on the eigenvalues and the spectral density, because there
are many more off--diagonal matrix elements than diagonal ones. This
is also clearly borne out in Noh's~\cite{Noh} and in our numerical
results. Moreover, there is no structural difference for the
off--diagonal matrix elements.  Hence, we are convinced that the
results of the present work carry over to models involving a
normalization of the kind~(\ref{eq0.1}). Our main reason to work with
the normalization in Noh's model is the fact that the limiting case
for infinitely long time series $T\to\infty$ can conveniently be
handled analytically.

We mention in passing that practitioners in banks and investment
companies commonly work with time series having a typical length of
250 days. These time series are often exponentially weighted before
the analysis. We do not employ such a weighting procedure, because we
are interested in the question, how a information about correlations of
longer time series can, to some extent and under proper conditions, be
estimated from shorter time series.

Furthermore, a comment regarding the branches in the model is in
order. An international portfolio is likely to contain globally
operating, large companies. Their activities directly affect various
branches, but they are also under different economical influences in
different countries.  In such cases, it is common to add a further
label ''country'' to the label ''branch'', denoted $b$ in our
case. Generally speaking, it is often necessary to go from a
one--factor model of the type discussed in the present article to a
two--factor or multi--factor model.  From the viewpoint of a
simulation, this refined correlation structure implies the need for
precise information about the weights which generalize our $p_b$.  On
the other hand, not much changes from the viewpoint of data analysis,
which was the main motivation of the present work.

\section{Noisy Spectral Density before Power Mapping}
\label{appA}

In general, the spectral density $\rho_T(\lambda)$ for a given length
$T$ of the time series can be written as the spectral average of the
joint probability density function $P_T(\lambda(T))$ for the
eigenvalues $\lambda_k(T), \ k=1,\ldots,K$ of the correlation matrix
$C(T)$,
\begin{eqnarray}
\rho_T(\lambda) = \int d[\lambda(T)] P_T(\lambda(T))
          \sum_{k=1}^K \delta\left(\lambda-\lambda_k(T)\right) \ ,
\label{eqA.1}
\end{eqnarray}
where $d[\lambda(T)]$ stands for the product of all differentials
$d\lambda_k(T)$. In our notation, we distinguish the argument
$\lambda$ of the density and the eigenvalues $\lambda_k(T)$ of the
correlation matrix which are always written with their argument
$T$. In particular, Eq.~(\ref{eqA.1}) is also valid for $T\to\infty$
and we have
\begin{eqnarray}
\rho_\infty(\lambda) = \int d[\lambda(\infty)] P_\infty(\lambda(\infty))
          \sum_{k=1}^K \delta\left(\lambda-\lambda_k(\infty)\right) \ . 
\label{eqA.2}
\end{eqnarray}
As a consequence of Eq.~(\ref{eq3.2}), the joint probability densities
for finite and infinite $T$ can, to leading order $1/\sqrt{T}$, be
related according to
\begin{eqnarray}
P_T(\lambda(T)) &=& \int d[\lambda(\infty)] P_\infty(\lambda(\infty))
                    \int d[a] \prod_{l=1}^K G(a_l,1)
                         \nonumber\\
                & & \quad \prod_{k=1}^K 
                         \delta\left(\lambda_k(T)-\lambda_k(\infty)
                                    -\frac{v_k}{\sqrt{T}}a_k\right) 
                                         \nonumber\\
                &=& \int d[\lambda(\infty)] P_\infty(\lambda(\infty))
                         \nonumber\\
                & & \quad 
          \prod_{k=1}^K G(\lambda_k(T)-\lambda_k(\infty),v_k^2/T) \ .
\label{eqA.3}
\end{eqnarray}
Here, $G(z,w^2)$ is the Gaussian depending on the variable $z$,
centered at zero, with variance $w^2$. We plug Eq.~(\ref{eqA.3}) into
Eq.~(\ref{eqA.1}), do the integrals over $\lambda(T)$ and find
\begin{eqnarray}
\rho_T(\lambda) &=& \int d[\lambda(\infty)] P_\infty(\lambda(\infty))
                    \sum_{k=1}^K G(\lambda-\lambda_k(\infty),v_k^2/T) 
                               \nonumber\\
                &=& \sum_{k=1}^K 
                    \int_{-\infty}^{+\infty} d\lambda^\prime
                    G(\lambda^\prime-\lambda,v_k^2/T)
                               \nonumber\\
                & & \quad
                    \int d[\lambda(\infty)] P_\infty(\lambda(\infty))
                    \delta\left(\lambda^\prime-\lambda_k(\infty)\right) \ .
\label{eqA.4}
\end{eqnarray}
In the second step, we inserted the integration over $\lambda^\prime$
using a $\delta$ function. Since we may assume that the joint
probability density $P_T(\lambda(\infty))$ is invariant under the
exchange of its arguments $\lambda_k(\infty)$, we can employ
Eq.~(\ref{eqA.2}) and integrate over the eigenvalues. We arrive at
\begin{eqnarray}
\rho_T(\lambda) = \int_{-\infty}^{+\infty} 
                  \overline{G}_T(\lambda^\prime-\lambda,v^2) 
                  \rho_\infty(\lambda^\prime) d\lambda^\prime \ ,
\label{eqA.5}
\end{eqnarray}
where we defined the average 
\begin{eqnarray}
\overline{G}_T(z,v^2) = \frac{1}{K} \sum_{k=1}^K G(z,v_k^2/T) 
\label{eqA.7}
\end{eqnarray}
over the $K$ Gaussians.  The parameters $v_k^2$ are functions of the
sizes $\kappa_b$ and the weights $p_b$.  Thus, to leading order
$1/\sqrt{T}$, we can obtain the spectral density for finite $T$ from
that for infinite $T$ by convoluting the latter with a superposition
of Gaussians.

Formula~(\ref{eqA.5}) is valid for every spectral density
$\rho_\infty(\lambda)$. We now apply it to the spectral
density~(\ref{eq3.1}) and obtain
\begin{eqnarray}
\rho_T(\lambda)
 &=& \sum_{b=1}^B (\kappa_b-1)
     \overline{G}_T\left(\lambda-\frac{1}{1+p_b},v^2\right)
                         \nonumber\\
 & & 
   + \sum_{b=1}^B 
     \delta\left(\lambda-\frac{1+\kappa_bp_b}{1+p_b}\right) 
   + \kappa \overline{G}_T\left(\lambda-1,v^2\right) \ .
\label{eqA.6}
\end{eqnarray} 
As the second term contains the large, widely spaced eigenvalues, we
neglect the smearing out there.  With this additional approximation,
the result~(\ref{eqA.6}) is valid to leading order $1/\sqrt{T}$. It
gives an analytical motivation for the estimate~(\ref{eq3.11}).

\section{Noisy Spectral Density after Power Mapping}
\label{appB}

The crucial effect of the power mapping is the preservation of the
block structure in $C^{(q)}(T)$ up to order $1/\sqrt{T}$.  This is
evident from Eqs.~(\ref{eq4.11}) and~(\ref{eq4.12}).  Only the
inclusion of the order $1/T^{q/2}$ destroys this block structure as
seen in Eq.~(\ref{eq4.13}). Hence, to understand the noisy spectral
density up to order $1/\sqrt{T}$, we can apply the methods
of~\ref{appA} individually to those $\kappa_b\times\kappa_b$ blocks
which contain the companies of one branch. A modification occurs for
the $\kappa\times\kappa$ block collecting the companies which do not
belong to a branch. According to Eq.~(\ref{eq4.11}), this block is up
to order $1/\sqrt{T}$ still a diagonal matrix, implying that the
eigenvalues equal the diagonal elements. Since we have $p_b=0$, they
are given by
$\lambda_k^{(q)}(T)=(C_{kk}(T))^q=1+qa_{kk}\sqrt{2/T}$. On the other
hand, employing a line of reasoning similar to the one in
Sec.~\ref{sec3.2}, the eigenvalues of the blocks corresponding to a
branch will have have the form
\begin{eqnarray}
\lambda_k^{(q)}(T) = \lambda_k^{(q)}(\infty) + 
                     \frac{v_k^{(q)}}{\sqrt{T}} a_k^{(q)}
\label{eqB.1}
\end{eqnarray}
where $a_k^{(q)}$ are independent Gaussian distributed variables with
zero mean and unit variance. The parameters $v_k^{(q)}$ result from a
superposition of $\kappa_b$ terms of order unity.  Thus, the Gaussian
smearing out will be stronger, roughly by a factor $\kappa_b$, for the
eigenvalues $\lambda_k^{(q)}(\infty)$ of the blocks corresponding to a
branch than for the eigenvalues $\lambda_k^{(q)}(\infty)=1$ which do
not belong to a branch. If we make the assumption that the sizes
$\kappa_b$ are large, we can neglect the smearing out of the latter
eigenvalues.  We emphasize that this is an additional approximation
which is not motivated by the asymptotic expansion in $T$. Under this
assumption, we obtain from Eq.~(\ref{eq4.14}) to leading order
$1/\sqrt{T}$ the estimate
\begin{eqnarray}
\rho_T^{(q)}(\lambda)
 &=& \sum_{b=1}^B (\kappa_b-1)
        \overline{G}_T\left(\lambda-
        \left(1-\left(\frac{p_b}{1+p_b}\right)^q\right),
                                  (v^{(q)})^2\right)
                         \nonumber\\
 & & 
   + \sum_{b=1}^B \delta\left(\lambda-
  \left(1+(\kappa_b-1)\left(\frac{p_b}{1+p_b}\right)^q\right)\right)
   + \kappa \delta\left(\lambda-1\right) \ .
\label{eqB.2}
\end{eqnarray}  
Since the second term involves the large eigenvalues outside the bulk
whose spacing is large compared to the smearing out, we may assume
that the corresponding $\delta$ functions are not affected, either. To
find an estimate to leading order $1/T^{q/2}$, we must apply the
methods of~\ref{appA} to the entire matrix, because the block
structure is destroyed. We replace Eq.~(\ref{eqB.1}) with
\begin{eqnarray}
\lambda_k^{(q)}(T) = \lambda_k^{(q)}(\infty) + 
                     \frac{v_k^{(q)}}{\sqrt{T}} a_k^{(q)} + 
                     \frac{\widetilde{v}_k^{(q)}}{T^{q/2}} 
                                \widetilde{a}_k^{(q)} \ .
\label{eqB.3}
\end{eqnarray} 
As we are only interested in a qualitative discussion, we make the
further assumption that the $\widetilde{a}_k^{(q)}$ are independent
Gaussian variables with zero mean and unit variance.  The smearing out
to order $1/T^{q/2}$ will not affect the first term of
Eq.~(\ref{eqB.2}), because it is already of order
$1/\sqrt{T}$. However, we cannot neglect it in the third term, because
the parameters $\widetilde{v}_k^{(q)}$ are now of order $K$.  We obtain
\begin{eqnarray}
\rho_T^{(q)}(\lambda)
 &=& \sum_{b=1}^B (\kappa_b-1)
        \overline{G}_T\left(\lambda-
        \left(1-\left(\frac{p_b}{1+p_b}\right)^q\right),
                                  (v^{(q)})^2\right)
                         \nonumber\\
 & & 
   + \sum_{b=1}^B \delta\left(\lambda-
  \left(1+(\kappa_b-1)\left(\frac{p_b}{1+p_b}\right)^q\right)\right)
                         \nonumber\\
 & & 
   + \kappa \overline{G}_{T^q}\left(\lambda-1,
                      (\widetilde{v}^{(q)})^2\right) \ .
\label{eqB.4}
\end{eqnarray}
Replacing each of the sums over the functions $\overline{G}_T$ and
$\overline{G}_{T^q}$ with one Gaussian, we arrive at the
estimate~(\ref{eq4.15}).


\section*{References}
\begin{thebibliography}{30}

\bibitem{PorTh}      E.J.~Elton and M.J.~Gruber, 
                  {\it Modern Portfolio Theory and 
                  Investment Analysis} 
                  (J.~Wiley, New York 1995).

\bibitem{Deutsch}    R.~Eller and H.P.~Deutsch, 
                  {\it Derivate und Interne Modelle, 
                  Modernes Risikomanagement} 
                  (Sch\"affer-P\"oschel Verlag, Stuttgart 1998).

\bibitem{RM}         J.~Longerstaey, A.~Zangari and S.~Howard, 
                  {\it Risk $Metrics^{{\rm TM}}$-Technical Document} 
                  (J.~P.~Morgan, New York 1996).

\bibitem{Lal}        L.~Laloux, P.~Cizeau, J.P.~Bouchaud and M.~Potters
                  Phys. Rev. Lett. {\bf 83}, 1467 (1999).

\bibitem{Pl}         V.~Plerou, P.~Gopikrishnan, B.~Rosenow, 
                  L.A.N.~Amaral and H.E.~Stanley,
                  Phys. Rev. Lett. {\bf 83}, 1471 (1999).

\bibitem{Mehta}      M.L.~Mehta, 
                  {\it Random Matrices}, 2nd edition,
                  (Academic Press, San Diego 1990).

\bibitem{Haake}      F.~Haake,
                  {\it Quantum Signatures of Chaos},
                  2nd edition,
                  (Springer Verlag, Berlin 2001).

\bibitem{GMGW}       T.~Guhr, A.~M\"uller--Groeling and
                  H.A.~Weidenm\"uller, 
                  Phys. Rep. {\bf 299}, 189 (1998).

\bibitem{Burda}      Z.~Burda, J.~Jurkiewicz, M.A.~Nowak, 
                  G.~Papp and I.~Zahed,
                  {\tt cond-mat/0103108};
                  {\tt cond-mat/0103108}.

\bibitem{Gop}        P.~Gopikrishnan, B.~Rosenow, V.~Plerou, 
                  L.A.N.~Amaral and H.E.~Stanley, 
                  {\tt cond-mat/0011145}.

\bibitem{Gu2}        V.~Plerou, P.~Gopikrishnan, B.~Rosenow, 
                  L.A.N.~Amaral, T.~Guhr and 
                  H.E.~Stanley,
                  {\tt cond-mat/0108023}.

\bibitem{Man}        R.N.~Mantegna, 
                  Eur. Phys. J. {\bf B11}, 193 (1999).

\bibitem{MS}         R.N.~Mantegna and H.E.~Stanley,
                  {\it An Introduction to Econophysics}
                  (Cambridge University Press, Cambridge, 2000)

\bibitem{BP}         J.P.~Bouchaud and M.~Potters, 
                  {\it Theory of Financial Risks}
                  (Cambridge University Press, Cambridge, 2000)

\bibitem{Voit}       J.~Voit, 
                  {\it The Statistical Mechanics of Financial 
                  Markets}
                  (Springer, Heidelberg, 2001)

\bibitem{PB}         W.~Paul and J.~Baschnagel,
                  {\it Stochastic Processes: From Physics to 
                  Finance}
                  (Springer, Berlin, 1999)

\bibitem{Stanley1}   P.~Gopikrishnan, V.~Plerou, L.A.~Amaral, M.~Meyer 
                  and H.E.~Stanley,  
                  Phys. Rev. {\bf E60}, 5305 (1999).

\bibitem{Stanley2}   P.~Gopikrishnan, V.~Plerou, L.A.~Amaral, M.~Meyer 
                  and H.E.~Stanley,  
                  Phys. Rev. {\bf E60}, 6519 (1999).

\bibitem{Noh}        J.D.~Noh,
                  Phys. Rev. {\bf E61}, 5981 (2000), 
                  {\tt cond-mat/9912076}.

\bibitem{Ross}       S.~Ross,
                  J. Econ. Theory {\bf 13}, 341 (1976).

\bibitem{Mar}        M.~Marsili, 
                  {\tt cond-mat/0003241}.

\bibitem{Kull}       L.~Kullmann, J.~Kert\'{e}sz and R.N.~Mantegna, 
                  {\tt cond-mat/0002238}.

\bibitem{Wu}         F.Y.~Wu, 
                  Rev. Mod. Phys. {\bf 54}, 235 (1982).

\bibitem{Krengel}    U.~Krengel, 
                  {\it Einf\"uhrung in die Wahrscheinlichkeitstheorie 
                  und Statistik} 
                  (Vieweg, Braunschweig 1991).

\bibitem{TWettig}    J.J.M.~Verbaarschot and T.~Wettig, 
                  Ann. Rev. Nucl. Part. Sci. {\bf 50} (2000) 343;
                  {\tt hep-ph/0003017}.

\bibitem{Dys}        F.J.~Dyson, 
                  Revista Mexicana de F\'{\i}sica {\bf 20}, 
                  231 (1971).

\bibitem{Seng}       A.M.~Sengupta and P.P.~Mitra, 
                  {\tt cond-mat/9709283}.

\bibitem{RD}         R.~Dahlhaus, private communication,
                  Oberwolfach, 2002.

\end{thebibliography}
\end{document}